\documentclass[10pt,journal]{IEEEtran}
\usepackage{amsmath,amsfonts}
\usepackage[english]{babel}
\usepackage{hyphenat}
\usepackage{algorithmic}
\usepackage{algorithm}
\usepackage{array}
\usepackage[caption=false,font=normalsize,labelfont=sf,textfont=sf]{subfig}
\usepackage{textcomp}
\usepackage{fancyref}
\usepackage{stfloats}
\usepackage{url}
\usepackage{verbatim}
\usepackage{graphics, graphicx}
\usepackage{epstopdf}
\usepackage{cite}
\usepackage{xcolor}
\newcommand{\bm}[1]{\mbox{\boldmath{$#1$}}}
\newcommand{\cscgdist}[2]{\sim \mathcal{CN} \left( #1, #2 \right)}

\usepackage{url}
\usepackage{hyperref}
\usepackage{soul} 
\usepackage{tikz}
\definecolor{gold}{rgb}{0.85,.66,0}
\definecolor{cian}{rgb}{.02,.7,.95}
\definecolor{dgreen}{rgb}{0,.4,0}

\begin{document}

\title{Reconfigurable Intelligent Surfaces-Enabled Intra-Cell Pilot Reuse in Massive MIMO Systems}

\author{José Carlos Marinello Filho, Taufik Abrão, Ekram Hossain, and Amine Mezghani

\thanks{This work was supported in part by the National Council for Scientific and Technological Development (CNPq) of Brazil under Grants 200988/2022-0, and 310681/2019-7 and in part by a Discovery Grant from the Natural Sciences and Engineering Research Council of Canada (NSERC).}

\thanks{J. C. Marinello F. is with Electrical Engineering Department, Federal University of Technology PR, Cornélio Procópio, PR, Brazil (e-mail: \nohyphens{jcmarinello@utfpr.edu.br}). He is currently a Visiting Researcher in the Department of Electrical and Computer Engineering at the University of Manitoba, Canada. }

\thanks{T. Abrão is with the Electrical Engineering Department, State University of Londrina, PR, Brazil (e-mail: taufik@uel.br).}

\thanks{Ekram Hossain and Amine Mezghani are with the Department of Electrical and Computer Engineering, University of Manitoba, Winnipeg, MB R3T 5V6, Canada
(e-mail: ekram.hossain@umanitoba.ca; amine.mezghani@umanitoba.ca).}
}




\maketitle

\begin{abstract}
Channel state information (CSI) estimation is a critical issue in the design of modern massive multiple-input multiple-output (mMIMO) networks. With the increasing number of users, assigning orthogonal pilots to everyone incurs a large overhead that strongly penalizes the spectral efficiency (SE) of the system. It becomes thus necessary to reuse pilots, giving rise to pilot contamination, a vital 
performance bottleneck of mMIMO networks. Reusing pilots among the users of the same cell is a very desirable operation condition from the perspective of reducing training overheads; however, the intra-cell pilot contamination might become even worse due to the users' proximity. Reconfigurable intelligent surfaces (RISs), which are capable of smartly controlling the wireless channel, can be leveraged to achieve intra-cell pilot reuse. In this paper, our main contribution is an RIS-aided approach for intra-cell pilot reuse and the corresponding channel estimation method. Relying upon the knowledge of only statistical CSI, we then optimize the RIS phase-shifts based on a manifold optimization framework {and the RIS positioning based on a deterministic approach}. The extensive numerical results  
highlight the remarkable performance improvements achieved by the proposed scheme (for both uplink and downlink transmissions) compared to other alternatives.


\end{abstract}


\section{Introduction}
\IEEEPARstart{T}{he} forthcoming sixth generation (6G) of telecommunication systems is envisioned to enable novel applications like augmented/virtual reality, holographic telepresence, autonomous transportation, and others \cite{Zhang19}. These novel services demand ever-increasing performances of the network in terms of spectral efficiency (SE) and energy efficiency. Furthermore, due to the upsurge in urban population and the explosive growth of internet-of-things (IoT) applications, the 6G networks will have to cope with an unprecedentedly high number of connected devices \cite{Nguyen22}. The reconfigurable intelligent surfaces (RISs), also known as intelligent reflecting surfaces (IRSs), can customize the wireless propagation environment by the operation of a large number of reconfigurable passive reflecting elements, which collectively are able to manipulate the impinging electromagnetic wave and reflect it according to an optimized radiation pattern, focusing the signal in an intended point \cite{DiRenzo20}. An improved channel response between the base station (BS) and user equipment (UE) is obtained as a benefit, as well as extended coverage, the capability of bypassing obstacles, and improved user fairness.

Massive multiple-input multiple-output (mMIMO) improves the 
{SE} of cellular networks by 
multiplexing a large number of UEs {with space-division multiple access (SDMA)} \cite{Emil18, Sanguinetti20}. It is a key technology of fifth generation (5G) communication, and also is regarded as a prominent candidate for the  6G physical layer standard \cite{Zhang19}. For an mMIMO system operating under time-division duplex (TDD) mode, the channel state information (CSI) can be acquired by the transmission of uplink (UL) pilot sequences exploiting the reciprocity between UL and downlink (DL) channels. As a result, the training overhead is proportional to the number of different orthogonal pilot sequences ($\tau_p$) used. As the number of UEs grows, assigning every UE an orthogonal pilot would significantly decrease the network SE, making it necessary to employ pilot reuse. However, pilot reuse gives rise to pilot contamination, which is known as an important performance bottleneck of mMIMO systems.

\subsection{{Literature Review}}

{Since its inception \cite{Marzetta10}, mMIMO systems have been evolving from a theoretical concept to a practical technology, becoming a key component of {the current 5G} standard \cite{Parkvall17}. The limited channel coherence blocks due to user mobility and channel delay spread makes the reuse of pilots necessary, which with the large number of BS antennas, gives rise to a beamformed interference known as pilot contamination \cite{Marzetta10, Hoydis13, Rusek13, Larsson14, Lu14, Bjornson15, Bjornson16, Marinello16, Marinello19}. Several different approaches have been proposed with the objective of mitigating this critical impairment of mMIMO systems, including time-shifted pilots and data transmission between different users \cite{Fernandes13, Jin16}, power allocation \cite{Chien18, Xu17, Herman22}, pilot assignment \cite{Xu17, Marinello16b, Marinello17, Herman22}, and cell-free (CF)-based schemes \cite{Polegre21}.}

In \cite{Emil18}, a promising method for mitigating pilot contamination is proposed, which takes advantage of the covariance matrices of the UEs' channel to improve channel estimation and combiner/precoder performances. The authors have shown that under the proposed approach, the asymptotic performance of mMIMO becomes unbounded; however, in the regime of a large but finite number of antennas, inter-cell pilot contamination still results in a significant performance loss. {The problem of exploiting spatial correlation of propagation channels to mitigate pilot contamination is revisited in \cite{Sanguinetti20} and extended to a CF system in \cite{Polegre21}.} An even more appealing and challenging scenario consists of reusing the pilots within the same cell. The higher number of UEs per pilot decreases the training overhead, increasing the pre-log factors of the SE computation. However, the intra-cell pilot contamination might be stronger due to the proximity of the UEs, requiring an effective way to deal with it to make intra-cell pilot reuse feasible. Recently, a rate-splitting multiple access (RSMA) approach has been proposed to mitigate intra-cell pilot contamination in \cite{Mishra23}. For DL transmission, the authors have proposed a framework integrating RSMA with TDD mMIMO, designed precoding strategies, and optimized power allocation to maximize different network utility functions. 

{In this paper, we show  that one or multiple RISs can be employed in a communication system to enable intra-cell pilot reuse. The RISs have been considered an enabling technology for 6G \cite{Zhang19}. By inducing phase shift to the impinging signal at each reflecting element, an RIS is able to modify the propagation channel between UEs and BS to benefit the communication process \cite{DiRenzo20, Liaskos18, Basar19}.  They are usually composed of thin and light material, making them convenient to be easily attached to the ceilings and walls in indoor environments, or deployed on the facade of buildings in outdoor environments.  Several papers have proposed approaches for joint optimizing the BS beamforming and RIS reflection configurations \cite{Wu19, Rehman21, Li23, Souto23}. Besides, RISs are employed in conjunction with mMIMO systems in \cite{Wang23} with the objective {of} serving the users in the service dead zone, \emph{i.e.}, in locations where the direct links between the BS and users are severely attenuated. The authors obtain closed-form expressions for achievable rates, optimize the RIS reflection coefficients, and characterize channel hardening and favorable propagation properties for the investigated scenario.}

Few works have addressed the pilot contamination mitigation problem with the aid of RIS. In \cite{Luo21}, the authors have shown that breaking TDD channel reciprocity using different random RIS configurations in UL and DL can mitigate the pilot contamination. However, channel reciprocity is indeed the main advantage of TDD mode, which enables mMIMO operation with reasonable training overheads without 
DL pilots for DL channel estimation. Requiring DL pilot transmission usually severely penalizes the SE of mMIMO systems since the length of the pilots scales with the number of BS antennas. Besides, for a {CF} scenario with single-antenna access points (APs) aided by a single RIS, the authors derive closed-form performance expressions considering pilot contamination in \cite{Shi22} and propose a fractional power control algorithm for the system. However, the phase shifts of the RIS elements are not optimized, being all kept constant as $\pi/4$.

\subsection{{Contributions}}

\begin{figure}[!htbp]
\centering
\includegraphics[width=3in]{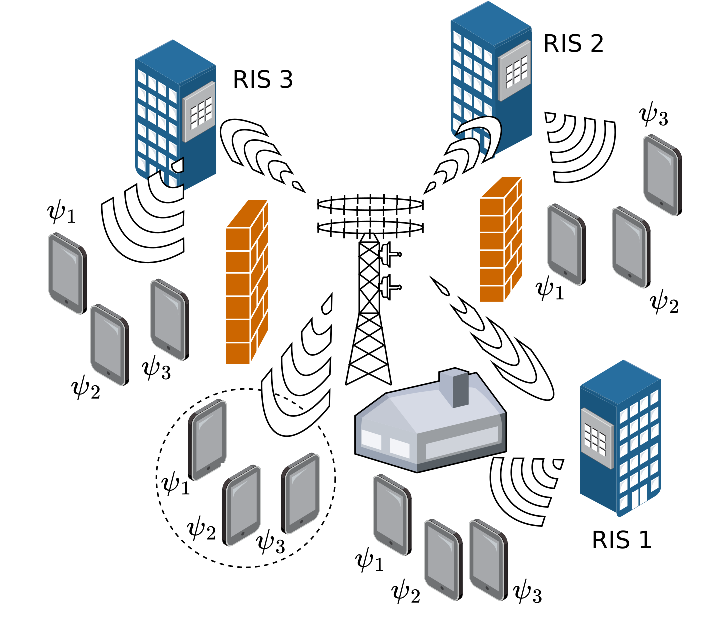}
\caption{{Multiuser mMIMO communication system assisted by multiple RISs, each deployed on the facades of buildings. The users in the dotted circle {area} are served without the aid of any RIS. They, as well as the users served by each RIS, share the same set of {pilot sequences $\{ \psi_1, \psi_2, \psi_3\}$} in our investigated scenario.} {Thus, each user sees interference of other users reusing the same pilot (IPR), as well as interference of users using other pilots (IOP).}}
\label{fig:spatial_dist}
\end{figure}

In this paper, we focus on the multi-user RIS-aided mMIMO operation in a realistic scenario where the channels have to be estimated from the transmission of scarce pilot sequences. 
{For scenarios with a high density of users, since it is desirable to keep the training overhead as low as possible aiming to obtain high SE,} we propose a method employing one or multiple RISs to allow multiple UEs {to share} 
a same pilot. Indeed, we show that {by} employing our proposed method, each additional RIS implemented in the cell allows a unit increment in the pilot reuse factor (PRF). This is equivalent to saying that up to $\tau_p$ UEs can connect to the network with the aid of a given RIS while they are assigned the pilots already in use in the cell. Thus, for example, while $\tau_p = 12$ would be required to serve 12 UEs in a cell with no pilot reuse, $\tau_p = 3$ would be enough to serve the same 12 UEs if employing a PRF of 4, \emph{i.e.}, reusing the same set of pilots 4 times among the UEs in the cell. To allow the PRF of 4, 3 RISs would be required in the cell. Note that 3 UEs can use the pilots without the aid of an RIS, while every other group of UEs has to be aided by an RIS to reuse the pilots. The scenario is illustrated in Fig. \ref{fig:spatial_dist}. The entire set of pilots is assigned to the UEs inside the dotted circle, which 
the BS {serves} without the aid of any RIS. Then, the same set of pilots is reused for the UEs served by each RIS, so that no pilot is reused among the UEs covered by the same RIS.

More specifically, our \emph{contributions} are as follows:
\begin{itemize}
    \item We propose an innovative use case for the RIS in realistic multi-user mMIMO scenarios, which allows multiple UEs to share the same pilot within the same cell.
    \item We formulate the system model of the proposed scenario and show how to obtain channel estimates, {required only for the overall channel}. Then, we propose an RIS phase-shift optimization method to improve SE performance. The proposed method only requires the knowledge of statistical CSI, simplifying the system operation.
    \item We also propose a deterministic approach to obtain optimized angular positions for the RIS deployment around the cell BS in order to further improve the performance of the intra-cell pilot reuse scheme. {The proposed approach depends only on BS parameters and has to be carried out {once,} only when installing the RISs.}
    \item Extensive numerical results are provided to corroborate that one or multiple RISs can be employed to enable this challenging but beneficial {RIS-aided mMIMO} operation scenario {for mitigating intra-cell pilot contamination}.
\end{itemize}

\subsection{{Organization}}

The remainder of this paper is organized as follows. The system model is described in Section \ref{sec:model}. The proposed 
{RIS-aided intra-cell pilot reuse} approach is presented in Section \ref{sec:RISoptz}. In Section \ref{sec:results}, the numerical results for the proposed method are presented. The main conclusions are offered in Section~\ref{sec:conclusion}.

\subsection{{Notations}}

Boldface lowercase $\mathbf{a}$ and uppercase $\mathbf{A}$ letters represent vectors and matrices, respectively. $[{\bf A}]_{i,j}$ represents the $i$th row and $j$th column element of matrix $\bf A$.  Calligraphic letters $\mathcal{A}$ represent finite sets. $\mathbf{I}_n$ denotes the identity matrix of size $n$, and $\mathbf{0}_n$ denotes the zero column vector of length $n$. $|\cdot|$ represents the magnitude operator when applied over a scalar, while representing the cardinality operator when applied over a set. $||\cdot||$ represents the Euclidean norm of a vector. $\{\cdot\}^*$, $\{\cdot\}^T$ and $\{\cdot\}^H$ denote, respectively, the complex conjugate operation, the transpose, and the complex conjugate transpose operators. ${\rm diag}(a_1, a_2, \ldots a_n)$ denotes a diagonal matrix whose diagonal entries are $a_1, a_2, \ldots a_n$. Besides, $\otimes$ denotes the Kronecker product while $\odot$ denotes the Hadamard product. ${\rm sinc}(\cdot)$ is the sinc function, ${\rm mod}(\cdot, \cdot)$ is the modulus operation, and $\lfloor \cdot \rfloor$ truncates the argument. Finally, $\mathbb{E}[\cdot]$ is the statistical expectation operator.

\section{System Model and Assumptions}\label{sec:model}

\subsection{Signal and Channel Models}
We consider a single-cell multi-user mMIMO system in which a BS with $M$ linearly disposed antennas serves $K$ single-antenna UEs. {The BS boresight is the $x$-axis in our scenario.} To improve the cell coverage, $R$ RISs are employed in certain strategic cell positions where the UE-BS direct channel link 
is severely attenuated. Each RIS employs $N$ reflecting elements, {arranged} in a rectangular array with $N_v$ rows and $N_h$ columns, such that $N = N_h \times N_v$. The reflection coefficient of each RIS element can be independently configured via an RIS controller, which is connected to the BS. Furthermore, we assume a quasi-static flat-fading channel model and that the RIS coefficients always have unitary amplitude, without loss of generality. The UL received baseband signal ${\bf y} \in \mathbb{C}^{M}$ at the BS can be written as
\begin{eqnarray}\label{eq:RxSignal}
    {\bf y} = \sum_{r=1}^{R} \sum_{k \in \mathcal{K}_r} \left({\bf H}_r {\rm diag}(\boldsymbol{\phi}_r) {\bf h}_{r,k} + {\bf h}^d_{k} \right) \sqrt{\rho_k} x_k +\nonumber\\ \sum_{k \in \mathcal{K}_0}{\bf h}^d_{k} \sqrt{\rho_k} x_k + {\bf n},
\end{eqnarray}
where ${\bf H}_r \in \mathbb{C}^{M\times N}$ is the channel matrix between BS and the $r$th RIS, $\bm{\phi}_r \in \mathbb{C}^{N}$ is a vector with the phase-shift coefficients of the $r$th RIS, ${\bf h}_{r,k} \in \mathbb{C}^{N\times 1}$ is the channel vector between the $r$th RIS and the $k$th UE, ${\bf h}^d_{k} \in \mathbb{C}^{M}$ is the direct channel vector between the BS and the $k$th UE, $\rho_k$ is the UL transmit power of the $k$th UE, $x_k$ with $\mathbb{E}[|x_k|^2]=1$ is the information symbol of the $k$th UE, and ${\bf n} \in \mathbb{C}^{M}$ is an additive white Gaussian noise (AWGN) vector following ${\bf n} \cscgdist{\mathbf{0}_M}{\mathbf{I}_M}$. Finally, we define $\mathcal{K}_0$ as the set of UEs not aided by any RIS, and $\mathcal{K}_r$ as the set of UEs served by the $r$th RIS. For simplicity, we assume that no UE is covered by multiple RISs\footnote{{If a certain UE is aided by multiple RISs, its performance would be significantly improved; however, the pilot used by this UE would not be available for other UEs served by these RISs. Therefore, a trade-off exists between performance improvement for one UE and  the number of served UEs.}}, \emph{i.e.}, 
\begin{equation}
    \mathcal{K}_r \cap \mathcal{K}_{r'} = \emptyset, \quad \forall r\neq r'.
\end{equation}

We assume a Rician fading environment for the BS-RIS channel since the RIS can be easily deployed in positions with good propagation conditions to the BS, usually with line-of-sight (LoS). Therefore, the matrices ${\bf H}_r$ are computed as follows:
\begin{equation}
    {\bf H}_r = \sqrt{\beta^{\textsc{br}}_r}\left[\sqrt{\frac{\alpha^{\textsc{br}}_r}{1+\alpha^{\textsc{br}}_r}} {\bf H}^{\textsc{LoS}}_r + \sqrt{\frac{1}{1+\alpha^{\textsc{br}}_r}} {\bf H}^{\textsc{NLoS}}_r \right],
\end{equation}
in which $\beta^{\textsc{br}}_r$ is the large-scale fading coefficient between the BS and the $r$th RIS, $\alpha^{\textsc{br}}_r$ is the Rician factor between the BS and the $r$th RIS, ${\bf H}^{\textsc{LoS}}_r$ is the deterministic LoS component of the channel, and ${\bf H}^{\textsc{NLoS}}_r$ is the random NLoS component of the channel, with each entry following an independent and identically distributed (i.i.d.) Rayleigh distribution, \emph{i.e.}, $\left[{\bf H}^{\textsc{NLoS}}_r\right]_{i,j}\cscgdist{0}{1}, \forall i,j$. The large-scale fading coefficient is given by
\begin{equation}
    \beta^{\textsc{br}}_r = \beta_0 \cdot (d^{\textsc{br}}_r)^{-\kappa_{\textsc{br}}},
\end{equation}
in which $\beta_0$ is the path loss at a reference distance of 1 m, $d^{\textsc{br}}_r$ is the distance between the BS and the $r$th RIS, and $\kappa_{\textsc{br}}$ is the path-loss exponent between BS and RIS. The deterministic LoS component of the channel is obtained following a geometric model given by
\begin{equation}
    {\bf H}^{\textsc{LoS}}_r = \bm{a}_{\textsc{b}}(\vartheta^{\textsc{b}}_r) \,\bm{a}_{\textsc{r}}(\vartheta^{\textsc{r}}_r,\varphi^{\textsc{r}}_r)^H,
\end{equation}
in which $\vartheta^{\textsc{b}}_r$ is the azimuthal angle of arrival (AoA) of the signal arriving at the BS from the $r$th RIS w.r.t. the array boresight, and $\vartheta^{\textsc{r}}_r$ and $\varphi^{\textsc{r}}_r$ are the azimuthal and elevation angle of departure (AoD) of the signal departing from the $r$th RIS to the BS with respect to the array boresight, respectively. Besides, we define the array response vector of the BS uniform linear array as
\begin{equation}\label{eq:ab}
    \bm{a}_{\textsc{b}}({\vartheta^{\textsc{b}}_r}) = \bm{a}\left(\frac{2 d_B}{\lambda} \sin{\vartheta^{\textsc{b}}_r}, M\right),
\end{equation}
and the array response vector of the RIS uniform planar array as
\small
\begin{equation}
    \bm{a}_{\textsc{r}}({\vartheta^{\textsc{r}}_r,\varphi^{\textsc{r}}_r}) = \bm{a}\left(\frac{2 d_R}{\lambda} \sin{\vartheta^{\textsc{r}}_r}\cos{\varphi^{\textsc{r}}_r}, N_h\right) \otimes \bm{a}\left(\frac{2 d_R}{\lambda} \sin{\varphi^{\textsc{r}}_r}, N_v\right),
\end{equation}
\normalsize
in which $d_B$ is the BS antenna separation, $d_R$ is the reflective element separation at the RIS, $\lambda$ is the carrier wavelength, and the standard array response vector {for generic inputs $\phi$ and $F$} is
\begin{equation}\label{eq:ar_std}
    \bm{a}(\phi, {F}) = [1, e^{-\imath \pi \phi}, e^{-\imath 2\pi \phi}, \ldots,  e^{-\imath \pi ({F}-1) \phi}]^T,
\end{equation}
with $\imath^2 = -1$.

We assume the channel vectors ${\bf h}_{r,k}$ and ${\bf h}^d_{k}$ being distributed as ${\bf h}_{r,k} \cscgdist{\mathbf{0}_N}{\mathbf{R}^{\textsc{ru}}_{r,k}}$ and ${\bf h}^d_{k}\cscgdist{\mathbf{0}_M}{\mathbf{R}^{\textsc{bu}}_{k}}$, in which $\mathbf{R}^{\textsc{ru}}_{r,k}$ is the spatial correlation matrix between the $r$th RIS and the $k$th UE, and ${\mathbf{R}^{\textsc{bu}}_{k}}$ is the spatial correlation matrix between the $k$th UE and the BS. The latter is generated following the exponential correlation model in \cite{Emil18, Emil17} as
\begin{equation}
    [{\mathbf{R}^{\textsc{bu}}_{k}}]_{m,n} = \beta^{\textsc{bu}}_k \zeta^{|n-m|} e^{\imath (n-m)\vartheta_k}
\end{equation}
in which $\beta^{\textsc{bu}}_k$ is the average large-scale fading coefficient of the $k$th UE w.r.t. the BS (direct link only), $\zeta \in [0,1]$ is the correlation factor between adjacent antennas, $\vartheta_k$ is the azimuthal AoA of the signal arriving at the BS from the $k$th UE w.r.t. the array boresight. The coefficient $\beta^{\textsc{bu}}_k$ is given by
\begin{equation}
    \beta^{\textsc{bu}}_k = \beta_0 \cdot (d^{\textsc{bu}}_k)^{-\kappa_{\textsc{bu}}},
\end{equation}
where $d^{\textsc{bu}}_k$ is the distance of the $k$th UE to the BS, and $\kappa_{\textsc{bu}}$ is the path loss exponent between BS and UEs.

For the $\mathbf{R}^{\textsc{ru}}_{r,k}$ spatial correlation matrices, we assume isotropic scattering in the half-space in front of the RIS \cite{Wang23, Emil21}. Thus, we have that $\mathbf{R}^{\textsc{ru}}_{r,k} = \beta^{\textsc{ru}}_{r,k} \mathbf{R}$, in which
\begin{equation}
    \beta^{\textsc{ru}}_{r,k} = \beta_0 \cdot (d^{\textsc{ru}}_{r,k})^{-\kappa_{\textsc{ru}}},
\end{equation}
being $d^{\textsc{ru}}_{r,k}$ the distance of the $k$th UE to the $r$th RIS, and $\kappa_{\textsc{ru}}$ the path loss exponent between RIS and UEs. The generic correlation matrix for the RIS 
is obtained as in \cite{Wang23, Emil21} as
\begin{equation}
    [{\mathbf{R}}]_{m,n} = {\rm sinc} \left( \frac{2||{\bf u}_m - {\bf u}_n||}{\lambda} \right),
\end{equation}
where ${\bf u}_{\ell}$ is the 3-dimensional location vector of the $\ell$th element of the RIS
\begin{equation}
    {\bf u}_{\ell} = [0, \,\, i(\ell)\, d_R, \,\,\ j(\ell) \,d_R]^T,
\end{equation}
where $i(\ell) = {\rm mod}(\ell-1,N_v)$ and $j(\ell) = \lfloor (\ell-1)/N_v\rfloor$ are
the horizontal and vertical indices of $\ell$-th RIS element.

In order to  describe how the channel estimates can be acquired and how the combiners/precoders can be designed, we first define the global set of UEs as
\begin{equation}
    \mathcal{K} = \mathcal{K}_0 \cup \mathcal{K}_1 \cup \ldots \mathcal{K}_R, 
\end{equation}
and $|\mathcal{K}| = K$ holds. Then, we define the overall channel of each UE $k \in \mathcal{K}$ as 
\begin{equation}\label{eq:ovr_ch}
  {\bf h}_k =
    \begin{cases}
      {\bf h}^d_k, & \text{if $k \in \mathcal{K}_0$, }\\
      {\bf H}_{r(k)} {\rm diag}(\boldsymbol{\phi}_{r(k)}) {\bf h}_{{r(k)},k} + {\bf h}^d_{k}, & \text{otherwise,}\\
    \end{cases}       
\end{equation}
in which $r(k)$ returns which RIS aids the $k$th UE, 
\emph{i.e.}, $r(k) = r' \Leftrightarrow k \in \mathcal{K}_{r'}$. We also define ${\bf R}_k = \mathbb{E}[{\bf h}_k {\bf h}_k^H]$. Then, eq.\eqref{eq:RxSignal} can be rewritten in a much simpler form as
\begin{eqnarray}\label{eq:RxSignal2}
    {\bf y} = \sum_{k \in \mathcal{K}} {\bf h}_{k} \sqrt{\rho_k} x_k + {\bf n}.
\end{eqnarray}

\subsection{Channel Estimation}\label{subsec:CSIest}

Channel estimation is performed at the BS after the UEs transmit UL pilots. The network makes available a number $\tau_p$ of orthogonal pilot signals $\{\bm{\psi}_1, \bm{\psi}_2, ... \bm{\psi}_{\tau_p}\}$, where $\bm{\psi}_{t} \in \mathbb{C}^{\tau_p}$ satisfies $ ||\bm{\psi}_t||^2=\tau_p$, $t \in \{1,2,...,\tau_p\}$. 
{The $K$ UEs then share the $\tau_p$ available pilots.} Denoting as $\tau_c$ the number of samples per channel coherence block, a fraction $\tau_p/\tau_c$ of them is spent with the pilot transmission, \emph{i.e.}, the training overhead is $\tau_p/\tau_c$. The best channel estimation quality is achieved by assigning an exclusive orthogonal pilot for each UE, which would demand $\tau_p = K$, resulting in a large overhead when serving many UEs. To avoid this, we resort to an {\it intra-cell pilot reuse scheme}, employing a PRF of $\varsigma$. This is equivalent to saying that each pilot is reused $\varsigma$ times within the cell, and therefore, we have $\tau_p = K/\varsigma$, remarkably decreasing the training overhead. 

Let the pilot $\bm{\psi}_{c(k)}$ out of the $\tau_p$ different pilot sequences be assigned to the $k$th UE, with $c(k)\in\{1,2,...,\tau_p \}$. The UEs assigned to the pilot $\bm{\psi}_t$ are represented in the set $\mathcal{S}_t = \{{\ell}: c({\ell}) = t\}$, while $|\mathcal{S}_t| \leq \varsigma$. Therefore, in the UL pilot transmission stage, the BS receives the signal ${\bf Y} \in \mathbb{C}^{M\times \tau_p}$ given by
\begin{eqnarray}\label{eq:RxPilot}
    {\bf Y} = \sum_{k \in \mathcal{K}} {\bf h}_{k} \sqrt{\rho_k} \bm{\psi}_{c(k)}^T + {\bf N},
\end{eqnarray}
where ${\bf N} \in \mathbb{C}^{M\times \tau_p}$ is a noise matrix such that $\left[{\bf N}\right]_{i,j}\cscgdist{0}{1}, \forall i,j$. The signal $\mathbf{Y}$ is then correlated with $\bm{\psi}_t$ at the BS:
\begin{equation}\label{eq:yt}
\begin{array}{l l}
    \mathbf{y}_t & = \mathbf{Y}\frac{\bm{\psi}_{t}^{*}}{||\bm{\psi}_{t}||^2} = \sum_{i \in \mathcal{S}_t}\sqrt{\rho_i} \mathbf{h}_i + \mathbf{n}_t,
\end{array}
\end{equation}
where $\mathbf{n}_t=\mathbf{N}\frac{\bm{\psi}_{t}^{*}}{||\bm{\psi}_{t}||^2}$ represents the effective noise and has distribution $\mathcal{CN }(0, \frac{1}{\tau_p}\mathbf{I}_M)$. To obtain the channel estimates $\widehat{{\bf h}}_k$, we employ the minimum mean squared error (MMSE) approach of \cite{Emil18, Sanguinetti20}, assuming $\rho_k = \rho$, $\forall k$, as follows:
\begin{eqnarray}\label{eq:mmse}
    \widehat{{\bf h}}_k = \frac{1}{\sqrt{\rho}} {\bf R}_k {\bf Q}^{-1}_{c(k)} \mathbf{y}_{c(k)},
\end{eqnarray}
where ${\bf Q}_{c(k)}$ is defined as
\begin{eqnarray}
    {\bf Q}_{c(k)} =\frac{1}{\rho} \mathbb{E}\left[{{\bf y}_{c(k)} {\bf y}_{c(k)}^H}\right] = \sum_{i \in \mathcal{S}_{c(k)}} {\bf R}_i + \frac{1}{\rho \tau_p}{\bf I}_M.
\end{eqnarray}
{Note that $\mathcal{S}_{c(k)}$ corresponds to the set $\mathcal{S}_t$ when $t=c(k)$, and is the set of UEs reusing the same pilot of UE $k$.}

The estimates $\widehat{{\bf h}}_k$ and the estimation error $\widetilde{{\bf h}}_k = {\bf h}_k - \widehat{{\bf h}}_k$ are independent random vectors distributed as $\widehat{{\bf h}}_k \cscgdist{\mathbf{0}_M}{\mathbf{\Phi}_{k}}$ and $\widetilde{{\bf h}}_k \cscgdist{\mathbf{0}_M}{\mathbf{R}_{k} - \mathbf{\Phi}_{k}}$, with $\mathbf{\Phi}_{k} = {\bf R}_k {\bf Q}^{-1}_{c(k)} {\bf R}_k$ \cite{Emil18}. Besides, it is important to note that, 
{for} the MMSE estimation in \eqref{eq:mmse}, the BS should know 
the spatial correlation matrices of the UEs in advance. It is discussed in \cite{Emil18, Emil17} how these matrices can be obtained in practical situations.

\subsection{Combiner Design}

In order to recover the data signal $x_k$ from $\bf y$ in \eqref{eq:RxSignal2}, the BS employs a linear combining vector ${\bf v}_k \in \mathbb{C}^M$, evaluating
\begin{eqnarray}
    \widehat{x}_k = {\bf v}_k^H {\bf y}.
\end{eqnarray}

The combining vector can be computed following different strategies. Under a maximum ratio (MR) approach, it is 
obtained as ${\bf v}^{\textsc{mr}}_k = \widehat{{\bf h}}_k$. On the other hand, following a zero-forcing (ZF) approach, it is necessary to first gather all channel estimates in the same matrix as follows:
\begin{equation}
    {\widehat{\bf H}} = [\widehat{{\bf h}}_1, \widehat{{\bf h}}_2, \ldots \widehat{{\bf h}}_K].
\end{equation}
Then, the ZF combiner of each user is obtained as
\begin{equation}
    {\bf v}^{\textsc{zf}}_k = \left[ {\widehat{\bf H}} \left( {\widehat{\bf H}}^H {\widehat{\bf H}} \right)^{-1}\right]_{:,k}.
\end{equation}

An MMSE combiner is also proposed in \cite{Emil18}. To compute it, it is necessary to first obtain the matrix $\bf Z$ as follows:
\begin{equation}
    {\bf Z} = \left[ \sum_{k=1}^K \left(\widehat{{\bf h}}_k \widehat{{\bf h}}_k^H + \mathbf{R}_{k} - \mathbf{\Phi}_{k} \right) + \frac{1}{\rho} {\bf I}_M \right]^{-1}.
\end{equation}
Then, ${\bf v}^{\textsc{mmse}}_k$ is simply computed as ${\bf v}^{\textsc{mmse}}_k = {\bf Z} \,\widehat{{\bf h}}_k$.

\subsection{Precoder Design}

{In the DL data transmission, the BS transmits the signal ${\bf s} \in \mathbb{C}^M$. This signal is given by}
\begin{equation}
    {\bf s} = \sum_{k \in \mathcal{K}} \sqrt{\rho_k} {\bf w}_k \varrho_k, 
\end{equation}
{in which ${\bf w}_k \in \mathbb{C}^M$ is the precoding vector associated with the $k$th UE, and $\varrho_k$ is the information symbol intended to this UE. It holds that $\mathbb{E} [||{\bf w}_k||^2] = 1$, and $\mathbb{E} [|\varrho_k|^2] = 1$. The signal received by the $k$th UE is}
\begin{equation}
    r_k = {\bf h}^H_k {\bf s} + \eta_k,
\end{equation}
{in which $\eta_k$ is the AWGN noise sample at the UE following $\eta_k \cscgdist{0}{1}$.}

{As discussed in \cite{Emil18}, the precoding vectors ${\bf w}_k$ can be designed following the MR, ZF, or MMSE criteria in the same way as the combining vectors ${\bf v}_k$ in the UL. One has just to ensure the unit norm condition, evaluating ${\bf w}_k = \frac{{\bf v}_k}{||{\bf v}_k||}$.}

\section{RIS-Aided Pilot Reuse, Optimization of RIS Phase-Shifts and RIS Position}\label{sec:RISoptz}

\subsection{RIS-Aided Pilot Reuse}

We introduce our proposed approach for employing the RIS to enable intra-cell pilot reuse. The main idea of our method is that each RIS can aid the communication of UEs using pilot sequences already in use in the cell, leveraging its reflection features. However, since the RIS configuration is unique for all UEs it serves, one must ensure that no pilot is reused between UEs aided by the same RIS, or between UEs aided by no RIS. 
Therefore, we assume that a prior scheduling algorithm has been carried out in such a way that the following condition holds:
\begin{equation}\label{eq:cond_sched}
    \mathcal{S}_{c(k)} \cap \mathcal{K}_{r(k)} = k, \quad \forall k \in \mathcal{K}.
\end{equation}
%
To meet the condition above, we arrive at the following relation between the PRF and the number of RISs in the cell: $\varsigma \leq R+1$.

Our proposed approach is based on the spatial correlation matrices of the overall channels in \eqref{eq:ovr_ch}. 
{Two main reasons justify the objective of relying only on statistical CSI:}
\begin{itemize}
    \item The statistical CSI remains valid for much larger time intervals than the actual CSI; thus we avoid in this way to require frequent reconfigurations of the RIS phase-shifts setup, which might be power consuming and require excessive hardware complexity \cite{Rech23};
    \item By not relying on actual CSI, we achieve a simple operation scenario for the RIS where only the overall channel has to be estimated, but the isolated ones have not. In this way, it decreases the necessity of employing active elements in the RIS, as well as the associated training overhead.
\end{itemize}

It is straightforward to see that ${\bf R}_k = {\bf R}^{\textsc{bu}}_k$ if $k \in \mathcal{K}_0$. Notwithstanding, we have to determine the spatial correlation matrices of UEs $k \in \mathcal{K}_r, r=1,2,\ldots R$. We do this by assuming that the BS-RIS channel matrices are constant and \emph{a priori} known by the BS. This assumption is motivated by the fact that the BS and the RIS are usually static, and therefore the associated channel coherence 
interval is much larger \cite{Wu21}. Furthermore, the RIS can be preferably deployed in a position with a favorable LoS component towards the BS. Motivated by these conditions, several recent papers have assumed the \emph{a priori} knowledge of the BS-RIS channels \cite{Wu21, You20}.

Then, for the UEs $k \in \mathcal{K}_r, r=1,2,\ldots R$, and writing $r(k) = r$ for ease of notation, we have
\begin{align}\label{eq:Rk}
    {\bf R}_k &= \mathbb{E}[{\bf h}_k {\bf h}_k^H],\nonumber\\
    &= \mathbb{E}[({\bf H}_r {\rm diag}(\boldsymbol{\phi}_r) {\bf h}_{r,k} + {\bf h}^d_{k})({\bf H}_r {\rm diag}(\boldsymbol{\phi}_r) {\bf h}_{r,k} + {\bf h}^d_{k})^H],\nonumber\\
    &= \mathbb{E}[{\bf H}_r {\rm diag}(\boldsymbol{\phi}_r) {\bf h}_{r,k} {\bf h}_{r,k}^H {\rm diag}(\boldsymbol{\phi}_r)^H {\bf H}_r^H] + {\bf R}^{\textsc{bu}}_k,
\end{align}
\normalsize
where from the second to the third line, we have used the fact that the direct BS-UE channel is independent of the RIS-aided channel, and with zero mean.

Then, treating the BS-RIS channel as a constant, as well as the RIS configuration, we can simplify \eqref{eq:Rk} as follows:
\begin{align}\label{eq:Rk2}
    {\bf R}_k &= \mathbb{E}[{\bf H}_r {\rm diag}(\boldsymbol{\phi}_r) {\bf h}_{r,k} {\bf h}_{r,k}^H {\rm diag}(\boldsymbol{\phi}_r)^H {\bf H}_r^H] + {\bf R}^{\textsc{bu}}_k,\nonumber\\
    &= {\bf H}_r {\rm diag}(\boldsymbol{\phi}_r) \mathbb{E}[{\bf h}_{r,k} {\bf h}_{r,k}^H] {\rm diag}(\boldsymbol{\phi}_r)^H {\bf H}_r^H + {\bf R}^{\textsc{bu}}_k,\nonumber\\
    &= {\bf H}_r {\rm diag}(\boldsymbol{\phi}_r) \mathbf{R}^{\textsc{ru}}_{r,k} {\rm diag}(\boldsymbol{\phi}_r)^H {\bf H}_r^H + {\bf R}^{\textsc{bu}}_k,\nonumber\\
    &= \beta^{\textsc{ru}}_{r,k} \, {\bf H}_r \, {\rm diag}(\boldsymbol{\phi}_r) {\bf R}\, {\rm diag}(\boldsymbol{\phi}_r)^H {\bf H}_r^H + {\bf R}^{\textsc{bu}}_k.
\end{align}
\normalsize

Eq. \eqref{eq:Rk2} reveals how the RIS can be employed to reconfigure the resultant statistical CSI of a given UE smartly. Thus, {one can leverage it} 
to optimize the RIS phase shifts. For example, as discussed in \cite{Emil18}, one can obtain orthogonal MMSE channel estimates between UEs $k$ and $k'$ whenever ${\bf R}_k {\bf R}_{k'} = {\bf 0}$. In this case, {one can say} 
that these UEs' correlation matrices have orthogonal support, and thus the pilot contamination vanishes when they share the same pilot and $M$ increases. However, seeking to optimize the RIS phase shifts under such target encounters several obstacles. First, the problem is manageable if 
{only two UEs are} sharing the same pilot; however, as this number grows to higher values of $\varsigma$, it would be necessary to consider every pair of UEs, with the problem assuming a combinatorial nature of increased complexity. Second, since the same RIS is used by several UEs, when the RIS phase-shifts are optimized to meet this condition for a UE using a certain pilot, it is unlikely that such condition will be met for other UEs too. Third, one must be careful that for the obtained solution of ${\bf R}_k {\bf R}_{k'} = {\bf 0}$,  ${\rm tr} ({\bf R}_k)$ and ${\rm tr} ({\bf R}_{k'})$ do not approach 0. Since ${\rm tr} ({\bf R}_k)$ can be seen as the average large-scale fading of UE $k$:
\begin{eqnarray}
    {\mathbb{E}\left[ ||{\bf h}_k||^2 \right] = \mathbb{E}\left[ {\bf h}_k^H {\bf h}_k \right] = \mathbb{E}\left[ {\rm tr}\left({\bf h}_k {\bf h}_k^H \right) \right] = {\rm tr}({\bf R}_k),}
\end{eqnarray}
it is desirable that it assumes higher values.

{Motivated by the above discussion, we first propose a decentralized optimization for the RIS phase-shifts in which the $r$th RIS optimizes its own phase-shifts aiming to maximize the average large-scale fading of the channels between its served UEs and the BS, according to eq. \eqref{eq:trRk}. Then, {in subsection \ref{subsec:positions},} we propose a method to obtain an optimized grid for the angular positions for the RIS 
{placement} in order to seek the condition of UEs served by different RISs having correlation matrices with orthogonal support, \emph{i.e.}, ${\bf R}_k {\bf R}_{k'} \approx {\bf 0}$, for $r(k) \neq r(k')$. By slightly adjusting\footnote{{For $M=128$ BS antennas with half-wavelength spacing, the highest distance between two angles in the optimized grid is $\approx 10$º. This means that, for an RIS deployed 100 m from the BS, the worst-case adjustment required for the RIS deployment position is of $\approx 8.8$ m.}} the RIS positions so that they fit into the optimized grid, we show how 
{one can remarkably reduce the interference between UEs served by different RISs.}}
\subsection{Optimizing the RIS Phase Shifts}\label{subsec:phase-shifts}

{First, we propose to optimize the RIS phase-shifts according to the following optimization problem:}
\begin{subequations}
\label{eq:trRk}
\begin{align}
P_0: \quad \underset{\bm{\phi}_r \in\, \mathbb{C}^N}{\mathrm{maximize}} \quad & {\rm tr}({\bf R}_k),\\
\label{eq:trRk_cstr}
\mathrm{subject\;to} \quad & |\phi_{r,i}| = 1; \; i = 1,2,\ldots,N.
\end{align}
\end{subequations}

However, the problem $P_0$ above has an undesirable characteristic in terms of multi-user RIS phase-shifts optimization. In particular, this is being UE-centric, \emph{i.e.}, if we optimize the RIS operation for a specific UE, it will not necessarily optimize the operation of the other UEs. Analyzing the objective function in \eqref{eq:trRk}, and the definition of ${\bf R}_k$ in \eqref{eq:Rk2}, we arrive at the following:
\begin{eqnarray}\label{eq:Rk3}
    {\rm tr}({\bf R}_k) = {\rm tr}\left(\beta^{\textsc{ru}}_{r,k} \, {\bf H}_r {\rm diag}(\boldsymbol{\phi}_r) {\bf R} {\rm diag}(\boldsymbol{\phi}_r)^H {\bf H}_r^H + {\bf R}^{\textsc{bu}}_k\right),\nonumber\\
    = \beta^{\textsc{ru}}_{r,k} \, {\rm tr}\left({\bf H}_r {\rm diag}(\boldsymbol{\phi}_r) {\bf R} {\rm diag}(\boldsymbol{\phi}_r)^H {\bf H}_r^H\right) + {\rm tr}\left({\bf R}^{\textsc{bu}}_k\right).
\end{eqnarray}
\normalsize

First, one can see that the second term in \eqref{eq:Rk3} does not depend on the RIS phase shifts. Thus, it can be neglected in the context of the optimization problem in \eqref{eq:trRk}. Then, the single dependence of ${\rm tr}({\bf R}_k)$ with the $k$th UE remains on the multiplicative factor $\beta^{\textsc{ru}}_{r,k}$, which can also be neglected when aiming to maximize ${\rm tr}({\bf R}_k)$ w.r.t. the RIS phase-shifts. Therefore, we obtain the following equivalent form for the optimization problem in \eqref{eq:trRk}:
\begin{subequations}
\label{eq:trRk4}
\begin{align}
P_1: \quad \underset{\bm{\phi}_r \in\, \mathbb{C}^N}{\mathrm{maximize}} \quad & {\rm tr}\left(\overline{\bf R}_r\right),\\
\label{eq:trRk4_cstr}
\mathrm{subject\;to} \quad & |\phi_{r,i}| = 1; \; i = 1,2,\ldots,N,
\end{align}
\end{subequations}
where 
\begin{eqnarray}\label{eq:Rk5}
    {\rm tr}\left(\overline{\bf R}_r\right) = {\rm tr}\left({\bf H}_r {\rm diag}(\boldsymbol{\phi}_r) {\bf R} {\rm diag}(\boldsymbol{\phi}_r)^H {\bf H}_r^H\right).
\end{eqnarray}

Problem $P_1$ in \eqref{eq:trRk4} can be seen as RIS-centric, \emph{i.e.}, since it is not specific for certain UE, its optimization leads to improved operating conditions for all UEs aided by the $r$th RIS simultaneously. Moreover, in \textbf{Appendix \ref{app:AppendA}}, we show 
that \eqref{eq:Rk5} can be rewritten in the following form:
\begin{equation}\label{eq:corrRIS}
    {\rm tr}\left(\overline{\bf R}_r\right) = \boldsymbol{\phi}_r^H {\bf D}_r \boldsymbol{\phi}_r,
\end{equation}
in which the matrix ${\bf D}_r$ is given by
\begin{equation}\label{eq:Dmat}
    {\bf D}_r = {\bf R}^* \odot [{\bf H}^H_r {\bf H}_r].
\end{equation}
In this way, we can rewrite the optimization problem\footnote{{If the UEs aided by the RIS have different correlation matrices ${\bf R}^{\textsc{ru}}_{r,k}$, the methodology proposed in this paper applies as well. One has just to replace the objective function in \eqref{eq:trRk} by $\sum_{k \in \mathcal{K}_r}{\rm tr}({\bf R}_k)$, and the $\bf R$ matrix in \eqref{eq:Dmat} by $\sum_{k \in \mathcal{K}_r}{\bf R}^{\textsc{ru}}_{r,k}$. {This scenario is also evaluated in the publicly shared simulation codes, as described in Section \ref{sec:results}}.}} in \eqref{eq:trRk4} as follows:
\begin{subequations}
\label{eq:trRk6}
\begin{align}
P_1: \quad \underset{\bm{\phi}_r \in\, \mathbb{C}^N}{\mathrm{maximize}} \quad & \boldsymbol{\phi}_r^H {\bf D}_r \boldsymbol{\phi}_r,\\
\label{eq:trRk6_cstr}
\mathrm{subject\;to} \quad & |\phi_{r,i}| = 1; \; i = 1,2,\ldots,N.
\end{align}
\end{subequations}

{In Problem \eqref{eq:trRk6}, one can see that if we did not have the constraints in \eqref{eq:trRk6_cstr}, the solution would be the eigenvalue of ${\bf D}_r$ corresponding to its maximum eigenvalue. However, the unit modulus constraints do not allow us to arrive at such a simple solution. Herein, we resort to a manifold optimization procedure \cite{Boumal_2023} to solve problem \eqref{eq:trRk6}. For this purpose, the Euclidean gradient of the objective function is given by}
\begin{equation}
    \nabla {\rm tr}\left(\overline{\bf R}_r\right) = 2 \, {\bf D}_r \boldsymbol{\phi}_r.
\end{equation}

{The choice for the manifold optimization procedure is motivated by the fact that the unit modulus constraints define a Riemannian manifold \cite{Yu19}. The geometry of Riemannian manifolds makes it possible to evaluate the gradients of cost functions, turning the optimization on manifolds locally analogous to that in Euclidean space. Thus, optimization techniques originally developed for Euclidean spaces have their counterparts on manifolds. {One can also employ other algorithms for resolving the optimization problem in \eqref{eq:trRk6}, such as the semidefinite relaxation (SDR) methods, since it is a quadratic programming problem with a concave objective function and non-convex constraints. However, it was shown in \cite{Yu19} that the manifold optimization method can usually achieve solutions without any performance loss in comparison with SDR.} In terms of computational complexity, the worst-case computational complexity when using conjugate gradient-based manifold optimization is $\mathcal{O}(N^{1.5})$ \cite{Yu19}, which is significantly less than when using {SDR} methods that entail a computational complexity of $\mathcal{O} ((N+1)^6)$. Besides, the availability of free solvers for most programming languages fairly simplifies the application of manifold optimization procedures \cite{Boumal_2023}.}

\subsection{Optimizing the RIS Positions}\label{subsec:positions}

{We present in this subsection a method to guide the choice of the deployment positions of the RISs within the cell. Our final result here is a set of optimized angles (notice that the distance to the BS has no influence on the result), so that if one deploys the RIS positions in the obtained regions, the interference between UEs aided by different RISs is remarkably reduced.} {Besides, the proposed method only depends on certain BS parameters, {such as} number and distance between antennas, carrier wavelength, and angular orientation of the antenna array. Therefore, it has to be carried out only a single time when installing the RISs, and remains valid while a LoS component exists in the BS-RIS link, which is usually the case.}

{Considering two different UEs $k$ and $k'$ in the cell with independent channels, the level of interference between them is proportional to}
\begin{align}\label{eq:crossProd}
    \mathbb{E}\left[ |{\bf h}_k^H {\bf h}_{k'} |^2 \right] &= \mathbb{E}\left[ {\rm tr} \left(({\bf h}_k^H {\bf h}_{k'})^H {\bf h}_k^H {\bf h}_{k'} \right) \right],\nonumber\\
    &= \mathbb{E}\left[ {\rm tr} \left( {\bf h}_k {\bf h}_k^H {\bf h}_{k'} {\bf h}_{k'}^H \right) \right],\nonumber\\
    &= {\rm tr} \left( \mathbb{E}\left[ {\bf h}_k {\bf h}_k^H {\bf h}_{k'} {\bf h}_{k'}^H \right] \right),\nonumber\\
    &= {\rm tr} \left( {\bf R}_k {\bf R}_{k'} \right).
\end{align}

{Assuming that the UEs are aided by different RISs, \emph{i.e.}, $r(k) \neq r(k')$, and denoting $r(k) = r$ and $r(k') = r'$ for ease of notation, we show in {\textbf{Appendix \ref{app:AppendB}}} that the result in \eqref{eq:crossProd} can be reduced if we have $\bm{a}_{\textsc{b}}(\vartheta^{\textsc{b}}_r)^H \bm{a}_{\textsc{b}}(\vartheta^{\textsc{b}}_{r'}) = 0$. Recall that $\vartheta^{\textsc{b}}_r$ is the {azimuthal} AoA at the BS of the signal coming from the $r$th RIS, which corresponds in our scenario to the angular position of this RIS. According to {\eqref{eq:ab}}, and applying the magnitude operator, we can rewrite it as}
\begin{align}\label{eq:prod_ar}
    |\bm{a}_{\textsc{b}}(\vartheta^{\textsc{b}}_r)^H \bm{a}_{\textsc{b}}(\vartheta^{\textsc{b}}_{r'})| &= \left| \sum_{m=0}^{M-1} e^{\imath 2 \pi \frac{d_B}{\lambda} m (\sin \vartheta^{\textsc{b}}_r - \sin \vartheta^{\textsc{b}}_{r'})} \right|,\nonumber\\
    &= \frac{|1-e^{\imath 2 \pi \frac{d_B}{\lambda} M (\sin \vartheta^{\textsc{b}}_r - \sin \vartheta^{\textsc{b}}_{r'})}|}{|1-e^{\imath 2 \pi \frac{d_B}{\lambda} (\sin \vartheta^{\textsc{b}}_r - \sin \vartheta^{\textsc{b}}_{r'})}|},\nonumber\\
    &= \frac{|\sin\left( \frac{\pi M d_B}{\lambda} (\sin \vartheta^{\textsc{b}}_r - \sin \vartheta^{\textsc{b}}_{r'})\right)|}{|\sin\left( \frac{\pi d_B}{\lambda} (\sin \vartheta^{\textsc{b}}_r - \sin \vartheta^{\textsc{b}}_{r'})\right)|}.
\end{align}

{Thus, we find the conditions required for making the result in \eqref{eq:prod_ar} equal to 0, which gives rise to a set of solutions for $\vartheta^{\textsc{b}}_r$. These solutions correspond thus to our obtained grid of angular positions for the RIS {placement}. The conditions are:}
\begin{equation}\label{eq:condition}
    \sin \vartheta^{\textsc{b}}_{r} = \sin \vartheta^{\textsc{b}}_{r'} + \ell \frac{\lambda}{M \,d_B}, \quad \ell \in \mathbb{Z} \; \& \; \ell \neq n M, n \in \mathbb{Z}.
\end{equation}

{Note that, the condition of $\ell$ being an integer non-multiple of $M$ (including 0) is required to null the numerator of \eqref{eq:prod_ar} without nulling its denominator.}
{It is also worth noting that, whenever the RISs of the cell are deployed in angular positions satisfying \eqref{eq:condition} for all pair of RISs, the interference between their aided UEs is significantly reduced. To gain further insights related to the allowed angular positions for the RIS deployment, we obtain in the following a grid of angular positions satisfying \eqref{eq:condition}.}

{We begin by admitting that $\vartheta^{\textsc{b}}_{r} = 0$ is a possible deployment region for the RIS deployment, since it is a very natural choice. This choice implies that any $\vartheta^{\textsc{b}}_{r} = \sin^{-1}\left( n \frac{\lambda}{d_B} \right)$, $n \in \mathbb{Z}$, be not a valid angle. For example, if one have $d_B = \lambda$, no other RIS could be deployed with $\vartheta^{\textsc{b}}_{r} \in [\frac{\pi}{2}, \pi, \frac{3 \pi}{2}]$. Fortunately, as $d_B = \lambda/2$ is usually deployed, one has just to ensure that: {\bf i)} if $\vartheta^{\textsc{b}}_{r} = \sin^{-1}\left( \ell \frac{2}{M} \right) = \phi$, $\ell \in \mathbb{Z}$, $0< |\ell| \leq \frac{M}{2}$, is deployed, no other RIS is deployed with $\vartheta^{\textsc{b}}_{r} = \pi - \phi$; and {\bf ii)} if $\vartheta^{\textsc{b}}_{r} = \pm \pi/2$ is deployed, no other RIS is deployed with $\vartheta^{\textsc{b}}_{r} = \mp \pi/2$.}

\begin{figure}[!t]
\centering
\includegraphics[width=3.5in]{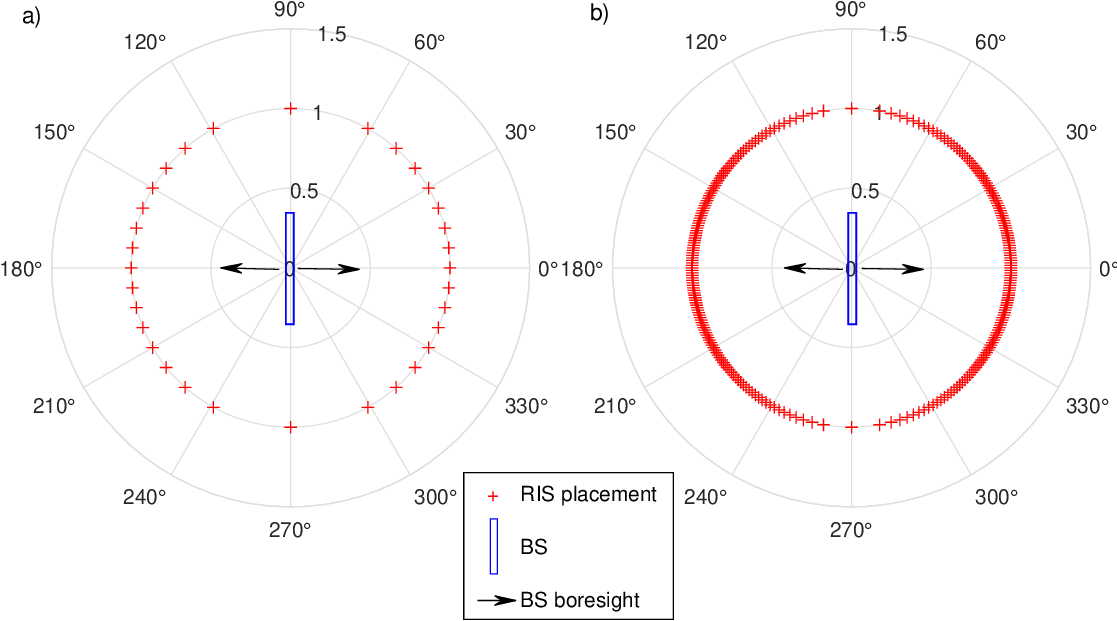}
\caption{Optimized grid of angular positions for the RIS deployment: a) $M=16$; and b) $M=128$ BS antennas. The BS boresight is the horizontal axis.}
\label{fig:optzAngs}
\end{figure}

{Therefore, to construct the grid with the optimized angles, we evaluate for the first quadrant $\vartheta^{\textsc{b}}_{r} = \sin^{-1}\left( \ell \frac{\lambda}{M \, d_B} \right)$, $\ell = 0, 1, \ldots, \frac{M \, d_B}{\lambda}$. We mirror this result with respect to the $x$-axis to the fourth quadrant, and mirror the result again with respect to the $y$-axis to the second and third quadrants. The obtained grid is depicted in Figure \ref{fig:optzAngs}, for $d_B = \lambda/2$ and $M = 16$ or 128 BS antennas. One can see that a total of $4 \frac{M \, d_B}{\lambda}$ angular positions are allowed in each scenario. Besides, as the angular resolution becomes lower near the $y$-axis, the worst-case resolution is $\Delta \phi_{\textsc{max}} = \frac{\pi}{2} - \sin^{-1}\left( 1 - \frac{\lambda}{M \, d_B} \right)$. In a typical scenario with $M = 128$ antennas and $d_B = \lambda/2$, $\Delta \phi_{\textsc{max}} = 0.177$ rad $\approx$ 10°; thus, deploying the RIS in a distance of 100 m from the BS, a maximum adjustment of $\approx 8.8$ m in the RIS position would be necessary in the worst-case. This result illustrates that slight adjustments in the RIS deployment positions are {generally} sufficient to meet the condition in \eqref{eq:condition}, which is an important guideline when choosing the positions to {install} the RISs in the cell. Furthermore, even if the condition in \eqref{eq:condition} cannot be exactly satisfied, it is very important to avoid making the equality hold with a certain $\ell$ integer multiple of $M$ (including 0), since in this way the interference between UEs aided by the different RISs would be maximized.}

\section{Numerical Results and Discussion}\label{sec:results}

\subsection{Simulation Set Up and Benchmarking}
We numerically evaluate the performance of the proposed RIS-aided intra-cell pilot reuse scheme in typical operation conditions. All simulation results of this paper can be reproduced using the codes at {GitHub}\footnote{\url{https://github.com/josecarlos-marinello/RIS-pilot-reuse}.}. {Table \ref{tab:simulation-parameters} summarizes the main simulation parameters}. {Note that although the number of pilots is kept low in the investigated system configurations ($\tau_p=1$ or $\tau_p=4$), the number of users does not, since in this section we investigate scenarios with up to 28 users by deploying PRF in the range $2\leq \varsigma \leq 7$. Hence, without no time/frequency splitting, some of these system setups can cover high-density UE scenarios.} {We assume a path-loss exponent of 2.3 in the BS-RIS and RIS-UE links due to the LoS existence and proximity with the UEs \cite{Wu21}. On the other hand, the path-loss exponent in the direct BS-UE link is considered 4.2 since the UEs aided by the RISs are assumed to be under severe channel attenuation.} {As benchmarks, we consider the mMIMO system without RIS (that we denote as \texttt{nr} in the figures herein) employing the techniques of \cite{Sanguinetti20}, which proposed signal processing schemes to resolve pilot contamination exploiting the natural spatial correlation of propagation channels. We also consider a scheme employing RIS with random phase-shifts for mitigating pilot contamination as proposed in \cite{Luo21} but employing the MMSE channel estimates described in \ref{subsec:CSIest}, {that we denote herein as \texttt{rps},} 
instead of the simpler least-squares (LS) estimates employed in \cite{Luo21}. On the other hand, the results with the RIS employing phase shifts obtained via the proposed manifold optimization approach are denoted by \texttt{mo}.} 

\begin{figure}[!t]
\centering
\includegraphics[width=3.5in]{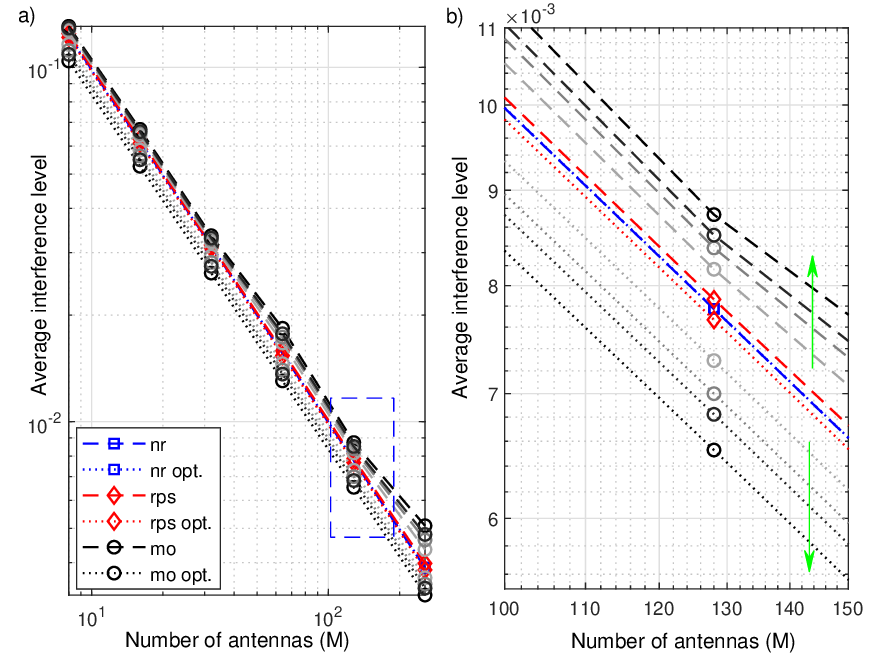}
\caption{Average normalized interference level between channel vectors of UEs aided by different RISs. We denote by "\texttt{opt}" when the RIS deployments positions are fitted to the optimized angular grid proposed here. Besides, the grey scale for the "\texttt{mo}" curves scales according to the {Rician} factor $\alpha_r^{\textsc{br}} \in [0, 3, 5, 10]$ dB. {The blue rectangle in a) is expanded in b), where the green arrows signalize the trend of the curves with the increasing Rician factor.}}
\label{fig:prodCorr}
\end{figure}

{We begin by evaluating the effectiveness of the proposed method for optimizing the angular positions for the RIS deployment. For this sake, we consider only two UEs, each one aided by a different RIS, deployed at a distance of $0.8 \, d_c$ from the BS. Each RIS deploys 16$\times$16 = 256 reflective elements (rows and columns, respectively) {arranged} in a rectangular grid. We first deploy the angular positions of the RIS at random, such that $\vartheta^{\textsc{b}}_{r} \sim \mathcal{U} [0, 2 \pi]$. Figure \ref{fig:prodCorr} depicts the average normalized interference levels for these UEs, defined as}
\begin{equation}\label{eq:avnormint}
    \varpi_{k,k'} = \frac{{\rm tr} \left( {\bf R}_k {\bf R}_{k'} \right)}{{\rm tr} \left( {\bf R}_{k} \right) {\rm tr} \left( {\bf R}_{k'} \right)},
\end{equation}
{with the increasing number of BS antennas. 
The figure compares the result with {\texttt{nr}, \texttt{rps}, and \texttt{mo} approaches}. Besides, for each scheme, we also present the result adjusting the original random angular positions of the RISs to the closest point in the optimized grid, denoted as \texttt{opt} in the figure. The procedure has no effect for the scenario without RIS, as expected, and very little influence with \texttt{rps}. This occurs basically because the performance improvement with \texttt{rps} is due to the received power increase {caused by} the RIS-aided link, but this takes almost the same effect in both the numerator and denominator of \eqref{eq:avnormint}. On the other hand, when employing the \texttt{mo} procedure to obtain the phase shifts, the average received signal power at the BS increases, which is seen as a higher interference for the other UE. Thus, the effect of smartly choosing the RIS deployment positions according to our proposed method becomes more visible, notably reducing the interference levels. Besides, the reduction becomes more significant as long as the LoS path in the BS-RIS link becomes predominant, represented by increasing the {Rician} factor in the figure. For example, for $M=128$ and $\alpha_r^{\textsc{br}} = 10$ dB,  the interference can be reduced from $8.733 \times 10^{-3}$ to $6.529 \times 10^{-3}$ ($\approx 25\%$).}

\begin{table}[!htbp]
\centering
\caption{Simulation parameters}
\label{tab:simulation-parameters}
\begin{tabular}{|c|c|}
\hline
\textbf{Parameter} & \textbf{Value}\\
\hline\hline
\multicolumn{2}{|c|}{\textit{\textbf{System}}} \\
\hline
Carrier frequency & $f_c = 3$ GHz\\
System bandwidth & $B = 10$ MHz\\
Number of antennas & $M \in \{ 8, 16, 32, 64, 128, 256\}$\\
Antennas spacing & $d_B = \frac{\lambda}{2} = 5$ cm\\
Pilot reuse factor & $\varsigma \in \{ 2, 3, 4, 5, 6, 7 \}$\\
Number of RISs & $R \in \{ 1, 2, 3, 4, 5, 6 \}$\\
RIS elements spacing & $d_R = \frac{\lambda}{2} = 5$ cm\\
RIS elements & $N = 256$\\
RIS elements disposition & $N_h \times N_v = 16 \times 16$\\
Number of users & $K \in \{ 4, 8, 12, 16, 20, 24, 28 \}$\\
Number of pilots & $\tau_p \in \{ 1, 4 \}$\\
Cell radius & $d_c = 150$ m\\
Transmit power & $\rho = 20$ dBm\\
RIS AoD & $\vartheta^{\textsc{r}}_{r} = \frac{\pi}{6}$\\
\hline
\multicolumn{2}{|c|}{\textit{\textbf{Channel}}}\\
\hline
Correlation factor & $\zeta = 0.5$\\
Path-loss at reference distance & $\beta_0 = -35.3$ dB\\
BS-RIS Rician factor & $\alpha_r^{\textsc{br}} = 5$ dB\\
BS-RIS path loss exponent & $\kappa_{\textsc{br}} = 2.30$\\
BS-RIS distance & $d^{\textsc{br}}_r = 0.8 \, d_c$\\
RIS-UE path loss exponent & $\kappa_{\textsc{ru}} = 2.30$\\
BS-UE path loss exponent & $\kappa_{\textsc{bu}} = 4.20$\\
Noise power spectral density & $-174$ dBm/Hz\\
Noise figure & $10$ dB\\
\hline
\multicolumn{2}{|c|}{\textit{\textbf{Monte-Carlo simulation}}}\\
\hline
Number of realizations & $S = 10^3$\\
\hline
\end{tabular}
\end{table}

\subsection{{UL Performance}}

\begin{figure}[!t]
\centering
\includegraphics[width=3.5in]{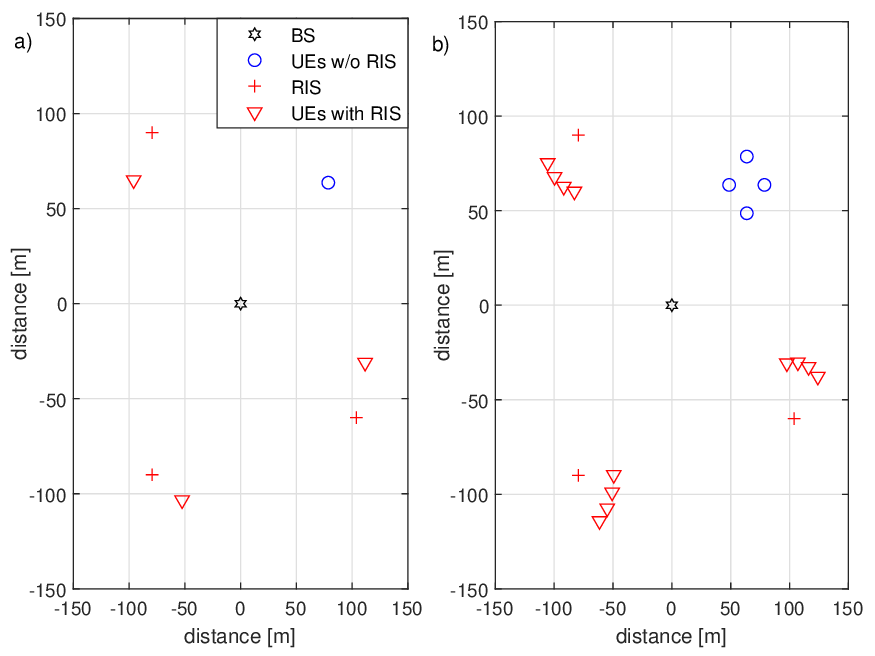}
\caption{Spatial topology of the network when  $\varsigma = 4$, and $R=3$: a) $K=4$, $\tau_p = 1$; and b) $K=16$, $\tau_p = 4$.}
\label{fig:spatial_dist2}
\end{figure}

{{We} analyze the system operating with $\varsigma = 4${,} $R=\varsigma -1 = 3$, {and spatial configuration} as illustrated in Fig. \ref{fig:spatial_dist2}. The closest UEs not aided by any RIS are uniformly distributed in a circle of radius $0.1 \, d_c$ centered in a distance of $0.6 \, d_c$ from the BS. The RISs are distant $0.8 \, d_c$ from the BS, while their aided UEs are distant $0.2 \, d_c$ from them. Besides, the center of the region containing the UEs without RIS as well as the $R$ RISs are initially regularly arranged in angles of $(n + \frac{1}{2}) \Delta \theta = (n + \frac{1}{2}) 2\pi/\varsigma$, with $n = 0, 1, \ldots \varsigma-1$. Then, the RIS positions are slightly adjusted according to our proposed method described in Section \ref{subsec:positions}. Finally, the AoD from every RIS to the BS is set to $\pi/6$, \emph{i.e.}, $\vartheta^{\textsc{r}}_r = \pi/6, \forall r$.} 

\begin{figure}[!t]
\centering
\includegraphics[width=3.5in]{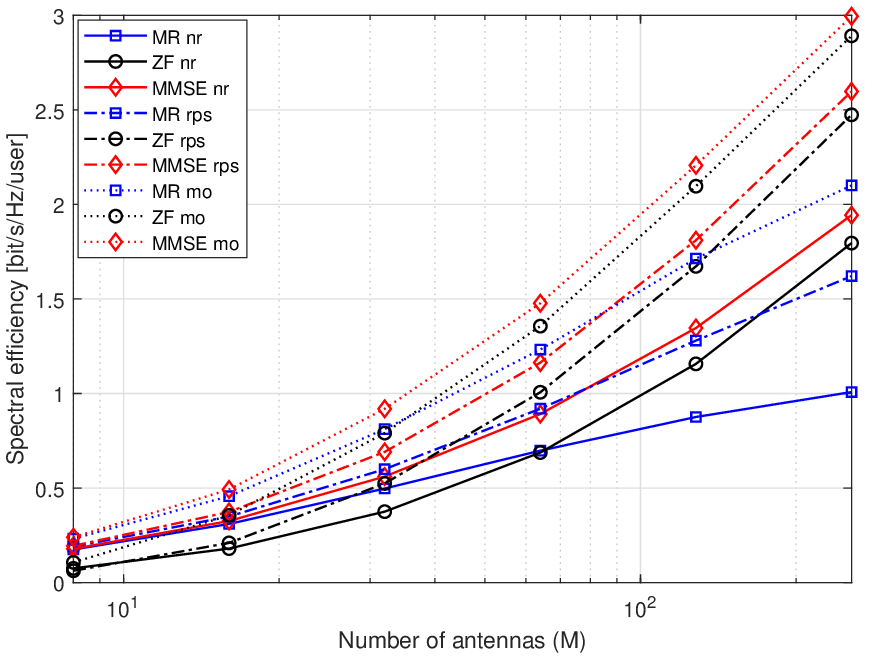}
\caption{UL SE with increasing $M$, when $R=3$, $N=256$, and $K=4$ UEs share the same pilot.}
\label{fig:SE_4ues_1pilot}
\end{figure}

Fig. \ref{fig:SE_4ues_1pilot} depicts the average UL SE performance with the increasing number of BS antennas $M$ {when $K=4$ UEs share a single pilot, with $\tau_p=1$, as illustrated in Fig.\ref{fig:spatial_dist2}.a.} 
Channel estimates are acquired through the MMSE approach of \eqref{eq:mmse}, while the MR, ZF, and MMSE combiners are employed. 
One can see a remarkable performance improvement {of} this last one compared to the others. For example, the average SE per user with $M=128$ antennas and MMSE combiner passes from 1.346 bit/s/Hz without RIS to 1.810 bit/s/Hz with \texttt{rps} ($\approx 34\%$ gain in SE) and 2.207 bit/s/Hz with \texttt{mo} ($\approx 64\%$ gain). Besides, one can also see that for any RIS configuration, the MMSE combiner always performs better than ZF, which performs better than MR, as expected.

\begin{figure}[!t]
\centering
\includegraphics[width=3.5in]{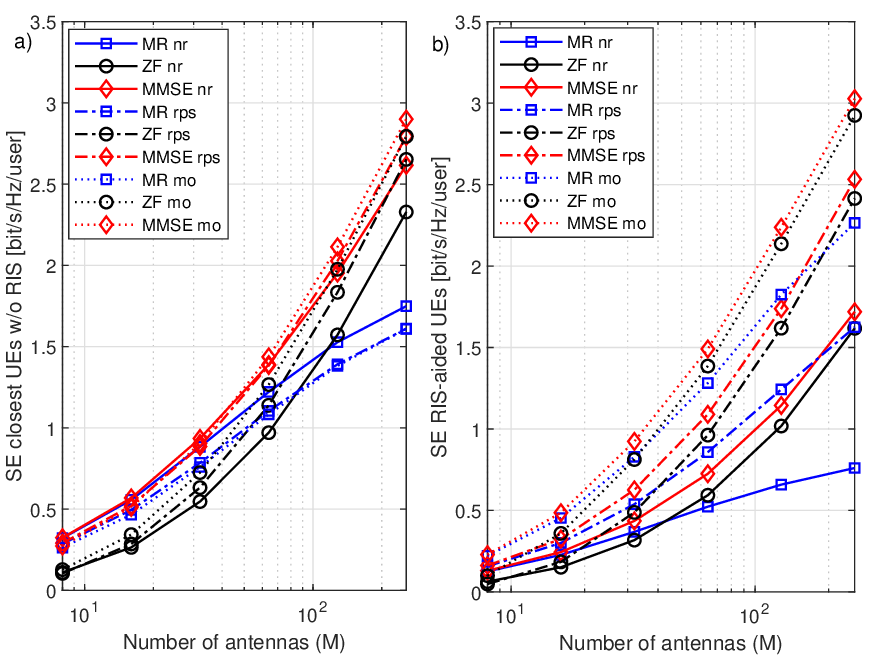}
\caption{UL SE with increasing $M$, when $R=3$, $N=256$, and $K=4$ UEs share the same pilot: a) closest UEs not aided by any RIS, b) farthest UEs aided by RIS.}
\label{fig:SE_4ues_1pilot_sp}
\end{figure}

Fig. \ref{fig:SE_4ues_1pilot_sp} depicts the UL SE performance of this same scenario, but showing in Fig. \ref{fig:SE_4ues_1pilot_sp}.a the average SE of the closest UEs which are not aided by any RIS, and in Fig. \ref{fig:SE_4ues_1pilot_sp}.b the average SE of the farthest RIS-aided UEs. It is worth noting that the intra-cell pilot reuse tends to be very harmful to the SE performance of the farthest UEs due to the severe intra-cell pilot contamination interference. However, the proposed scheme leverages the RISs to circumvent this scenario, optimizing the RIS reflection {and position} and improving the propagation conditions of such UEs. While the performance of the closest UEs does not suffer significant changes, the farthest UEs' SE employing MMSE combiner and $M=128$ antennas increases from 1.144 bit/s/Hz without RIS to 1.738 bit/s/Hz ($\approx 52\%$) with \texttt{rps} and 2.238 bit/s/Hz ($\approx 96\%$ gain in SE) with \texttt{mo}. As the SE of the closest UEs in this scenario is 2.114 bit/s/Hz, the proposed scheme is able to provide much more uniform SE performances for the UEs in the cell.

\begin{figure}[!t]
\centering
\includegraphics[width=3.5in]{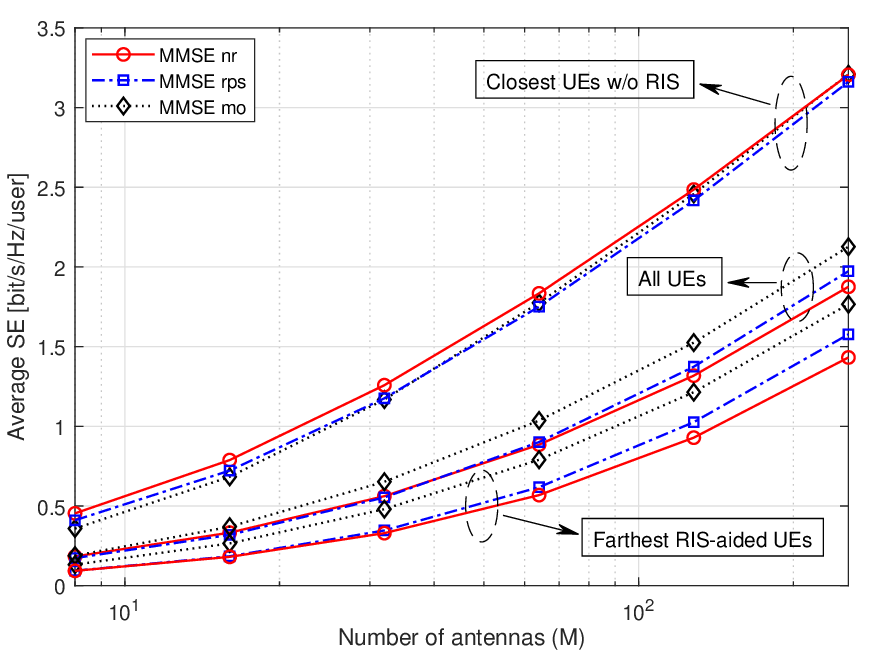}
\caption{UL SE with increasing $M$, when $R=3$, $N=256$, and $K=16$ UEs share $\tau_p=4$ pilots.}
\label{fig:SE_16ues_4pilot}
\end{figure}

Since the MMSE {scheme} achieves the best performance, only this approach is evaluated hereafter {in this subsection}. Fig. \ref{fig:SE_16ues_4pilot} depicts the average SE performance with the increasing number of BS antennas, considering $R=3$, $N=256$, and $K=16$ UEs sharing $\tau_p=4$ pilots {(configuration scenario in Fig. \ref{fig:spatial_dist2}.b)}. It shows the average performances taking into account only the closest UEs not aided by any RIS, only the farthest RIS-aided UEs, and all UEs as well. One can see that the SE performance of the closest UEs 
{does not suffer significant changes} employing the proposed approach, while the {average SE per user} performance of the farthest UEs is improved {regarding \texttt{nr} and \texttt{rps} approaches}. This occurs since the average channel gain of the RIS-aided UEs increases, 
{and} the resultant effect is {an} improvement in the SE when averaged between all UEs. The farthest UEs' SE employing $M=128$ antennas increases from 0.929 bit/s/Hz without RIS to 1.026 bit/s/Hz ($\approx 10\%$) with \texttt{rps} and 1.214 bit/s/Hz ($\approx 31\%$ gain in SE) with \texttt{mo} {method}. {When averaging between all UEs, the performance also increases $\approx 16 \%$ with \texttt{mo} in comparison with no RIS {scheme}.}

\begin{figure}[!t]
\centering
\includegraphics[width=3.5in]{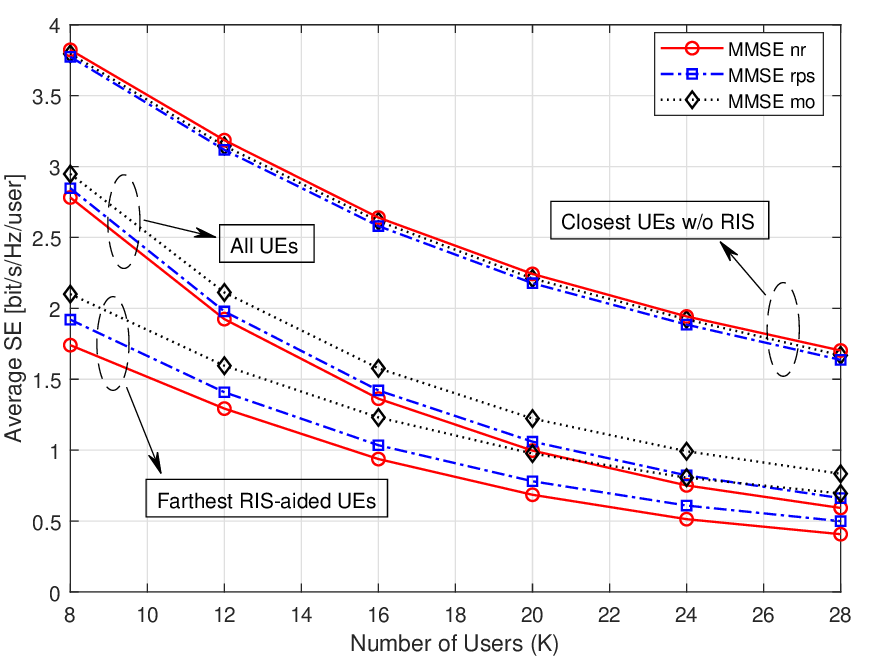}
\caption{UL SE with increasing {$K$, $R$, and $\varsigma$,} such that $K=\varsigma \, \tau_p = (R+1)\, \tau_p$, when $\tau_p=4$, $N=256$, and $M=128$ antennas at the BS.}
\label{fig:SE_4pilots_varK}
\end{figure}

{Then, we keep fixed the number of pilots as $\tau_p = 4$ and the number of BS antennas as $M=128$, and let the PRF $\varsigma$ increase together with the number of RISs and UEs, such that $\varsigma = R+1 = K/\tau_p$ holds. Fig. \ref{fig:SE_4pilots_varK} depicts how the UL SE is affected by such aggressive intra-cell pilot-reuse scenarios, while showing that our proposed methodology effectively leverages the RISs to improve performance in these challenging conditions. One can see an almost linear increase of $\approx 0.3$ bit/s/Hz per UE achieved with \texttt{mo} in comparison with no RIS {strategy} when averaging between the farthest RIS-aided UEs. As such UEs are a fraction of $\frac{\varsigma-1}{\varsigma}$ of the total number of UEs, the UL SE gain when averaging between all UEs starts from $\approx 0.16$ bit/s/Hz and gradually converges to the same increase of $\approx 0.3$ bit/s/Hz {as $K$ increases}. Besides, if one fixes a target UL SE performance of 1 bit/s/Hz, the number of UEs can be increased from 20 to 24 when employing our proposed methodology, with neither increase in the training overhead, nor significant increases in power consumption or processing complexity.}

{Finally, for the same UL scenario of Fig. \ref{fig:SE_4pilots_varK} and employing MMSE combiner, {we characterize} how the desired signal and interference terms behave with the increasing $\varsigma$, $R$, and $K$. For this sake, we recall from \cite{Emil18} that the UL signal-to-interference-plus-noise ratio (SINR) in the investigated scenario with uniform power allocation $\rho_k = \rho$, $\forall k$, is given by}
\begin{eqnarray}\label{eq:sinr}
    \gamma^{\textsc{ul}}_k = \frac{ |{\bf v}_k^H {\bf h}_k|^2}{\sum\limits_{\substack{k' \in \mathcal{K} \\ k'\neq k}} |{\bf v}_{k}^H {\bf h}_{k'}|^2 + \sum\limits_{k' \in \mathcal{K}} {\bf v}_k^H \left({\bf R}_k - \bm{\Phi}_k \right) {\bf v}_k + \frac{||{\bf v}_k||^2}{\rho}}.
\end{eqnarray}

\begin{figure}[!t]
\centering
\includegraphics[width=3.5in]{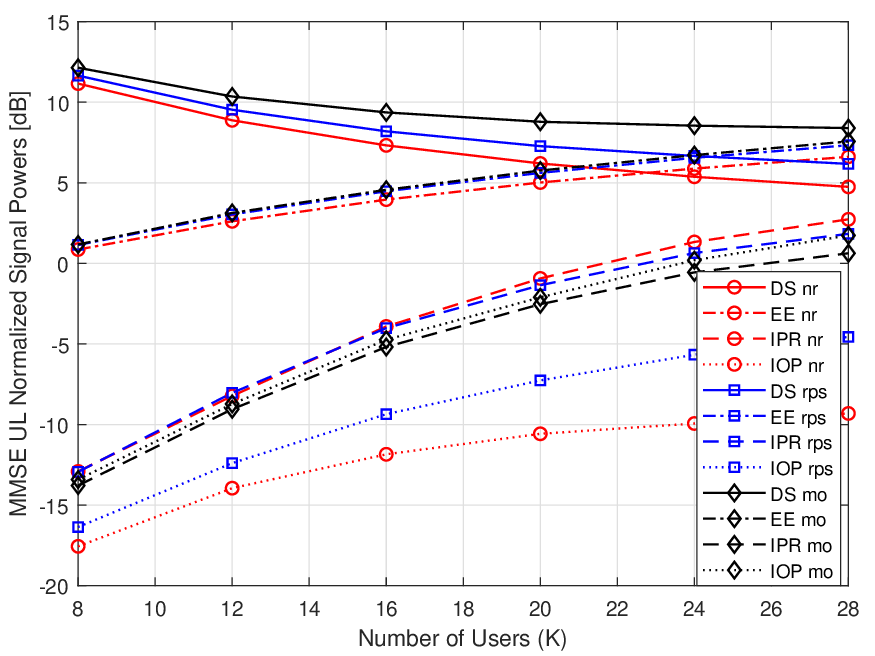}
\caption{UL normalized powers with MMSE combiner and increasing {$K$, $R$, and $\varsigma$,} such that $K=\varsigma \tau_p = (R+1) \tau_p$, when $\tau_p=4$, $N=256$, and $M=128$ antennas at the BS.}
\label{fig:Powers_4pilots_varK}
\end{figure}

{While the numerator of \eqref{eq:sinr} can be seen as the UL desired signal (DS) power, the first term of the denominator can be seen as the multi-user interference, the second one as a performance decrease due to the channel estimation error (EE), and the last one as the noise power at the combiner output. To evaluate how these terms behave with the increasing PRF, we normalize all of them with respect to the noise power at the combiner output. Besides, we decompose the multi-user interference in two terms: a first one regarding the UEs sharing the same pilot, {which} we denote as interference due to pilot-reuse (IPR), and a second one regarding the other pilots, {which} we denote as interference due to other pilots (IOP). Fig. \ref{fig:Powers_4pilots_varK} shows how the normalized powers of these terms vary with the increasing number of $K$, $R$, and $\varsigma$. One can see that our proposed method has the positive effect of significantly increasing the DS, while reducing the IPR. However, it {substantially} increases the EE, and significantly increases the IOP, although this last one still remains {at} significantly lower levels than the DS and EE. Since the improvement in the DS is much more significant than the increase in the EE, the final result is of improving the SINRs, which have been mapped into the SE gains of the previous figures.}

\subsection{{DL Performance}}

{We evaluate in this subsection the SE performance of our proposed RIS-aided intra-cell pilot reuse scheme for DL communication. In addition to the previous benchmarks, we also consider the scheme of \cite{Mishra23}, which proposed an RSMA-based method for mitigating intra-cell pilot contamination in the DL of mMIMO systems. The scheme of \cite{Mishra23} allows the $K$ UEs to communicate using the same pilot by employing RSMA, whereas it employs a weighted MR precoder for the common stream, and MR precoder for the private streams. Power allocation policies are proposed according to different metrics, among which we have chosen the MaxSum-SE one since we are evaluating the average SE per user. In this case, the authors propose a power allocation policy {that} exhaustively searches (in steps of 5\%) the best fraction of power allocated to the common stream, while the remaining power is uniformly distributed for the UEs' private streams. Besides, the MR common precoder {weights} are optimized aiming to maximize the minimum common SINR at the UEs employing CVX, a package to solve disciplined convex programs in Matlab \cite{cvx, gb08}. For a fair comparison with our scheme, we also adapted the method of \cite{Mishra23} to employ the MMSE precoder in the private streams, achieving thus better performances.} 

\begin{figure}[!t]
\centering
\includegraphics[width=3.5in]{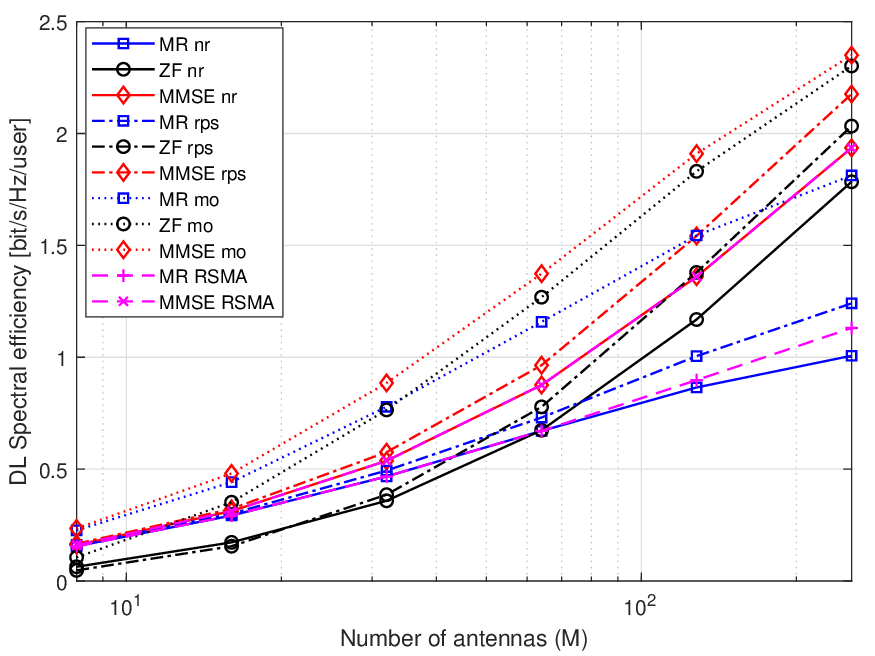}
\caption{DL SE with increasing $M$, when $R=3$, $N=256$, and $K=4$ UEs share the same pilot.}
\label{fig:SE_4ues_1pilot_DL}
\end{figure}

{Fig. \ref{fig:SE_4ues_1pilot_DL} evaluates the DL SE performance in the same scenario of Fig. \ref{fig:SE_4ues_1pilot}}. One can again see a remarkable performance improvement of our proposed method compared to the others, {similarly as has been seen in UL}. This is justified since our RIS configuration optimization approach aims to maximize the average channel gain of the UEs, which is the same for both UL {and} DL. Therefore, both communication links are simultaneously improved. For example, the average DL SE per user with $M=128$ antennas and MMSE precoder passes from 1.360 bit/s/Hz without RIS to 1.544 bit/s/Hz with \texttt{rps} ($\approx 14\%$ gain in SE) and to 1.910 bit/s/Hz with \texttt{mo} ($\approx 41\%$ gain). Again, the MMSE approach performs better than the other ones for all investigated RIS configurations and/or the absence of RIS. {Besides, our proposed scheme performs significantly better than the RSMA scheme proposed in \cite{Mishra23}. One can see that the RSMA employing MR precoder in the private streams performs slightly better than MR \texttt{nr}, which takes advantage of spatial correlation matrices to mitigate pilot contamination as proposed in \cite{Sanguinetti20}. This occurs since when using MR precoding, the effective channels are not sufficiently orthogonal; thus, it becomes optimal to allocate some power in the common stream and take advantage of the correlated channels, justifying the RSMA performance improvement, as shown in \cite{Mishra23}. However, when using MMSE precoder in the private streams, the effective channels achieve a good orthogonality condition, in such a way that the optimal alternative is to allocate all power in the private streams, making RSMA converge to SDMA. It is worth noting that the good orthogonality condition occurs even with all UEs using the same pilot, since the MMSE channel estimation and precoding of \cite{Emil18, Sanguinetti20} takes advantage of spatial correlation matrices for this purpose. On the other hand, our proposed method further boosts this condition by optimizing the RIS phase-shifts to increase the average channel gain and the RIS deployment positions to reduce interference between UEs aided by different RIS, being the most effective option for enabling {intensive} intra-cell pilot reuse.}

\begin{figure}[!t]
\centering
\includegraphics[width=3.5in]{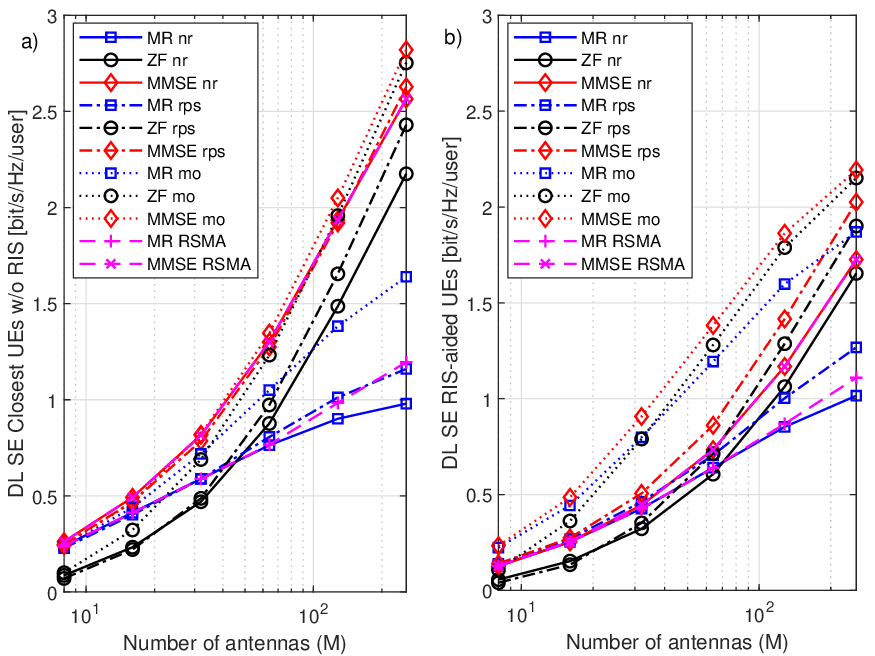}
\caption{DL SE with increasing $M$, when $R=3$, $N=256$, and $K=4$ UEs share the same pilot: a) closest UEs not aided by any RIS, b) farthest UEs aided by RIS.}
\label{fig:SE_4ues_1pilot_sp_DL}
\end{figure}

{Furthermore, Fig. \ref{fig:SE_4ues_1pilot_sp_DL} presents the same analysis of Fig. \ref{fig:SE_4ues_1pilot_sp}, but considering now the forward link. {Similar to what we have observed in the case of UL transmission, one can see that the performances of distant UEs are significantly improved by the use of the RISs, mainly when implementing our proposed optimization for the RIS reflection coefficients, while the performance of the closer UEs are almost the same (except for the MR precoding that is remarkably improved by our method).} The DL SE of the farthest UE, when employing MMSE precoder and $M=128$ antennas, increases from 1.168 bit/s/Hz without RIS to 1.416 bit/s/Hz ($\approx 21\%$) with \texttt{rps} and 1.863 bit/s/Hz ($\approx 60\%$ gain in SE) with \texttt{mo}. This last DL SE performance becomes closer to that achieved by the closest UEs, which is 2.045 bit/s/Hz.} {Besides, one can see again a remarkable performance improvement of our method {compared to} RSMA employed for intra-cell pilot reuse.}

\section{Conclusion}\label{sec:conclusion}

We have proposed an RIS-aided intra-cell pilot reuse approach for 6G scenarios. Since such systems are envisioned to serve a large number of users and assigning orthogonal pilots to everyone would incur large training overheads, the proposed method is a promising approach for achieving high spectral efficiency performance. We have proposed an optimization approach for the RIS phase-shifts able to reconfigure the resultant statistical CSI of the RIS-aided UEs, {and {also} for the RIS deployment positions in order to reduce interference between UEs aided by different RISs,} providing them a higher average channel gain while mitigating the intra-cell pilot contamination. As a result, the proposed method can assure good SE performance for the UEs without requiring a significant increase in complexity or energy consumption, {due} to decreased training overheads. {Possible research extensions of this work involve developing clustering algorithms for the RIS-UEs association, as well as extending the methodology for near-field communication between RIS and UEs.}



\appendices
\section{Proof of the Equation \eqref{eq:corrRIS}}\label{app:AppendA}
In order to arrive at \eqref{eq:corrRIS}, we first observe that ${\bf H}_r {\rm diag}(\boldsymbol{\phi}_r)$ has a block structure as follows
\begin{equation}
    {\bf H}_r {\rm diag}(\boldsymbol{\phi}_r) = \left[[{\bf H}_r]_{:,1} {\phi}_{r,1}, \ldots [{\bf H}_r]_{:,N} {\phi}_{r,N} \right].
\end{equation}
Then, we can write
\begin{eqnarray}\label{eq:Rr}
    \overline{\bf R}_r &=& {\bf H}_r {\rm diag}(\boldsymbol{\phi}_r) {\bf R} {\rm diag}(\boldsymbol{\phi}_r)^H {\bf H}_r^H,\nonumber\\
    &=& \sum_{i=1}^N \sum_{j=1}^N [{\bf H}_r]_{:,i} {\phi}_{r,i}\, [{\bf R}]_{i,j}\, [{\bf H}_r]^H_{:,j} {\phi}^*_{r,j},\nonumber\\
    &=& \sum_{i=1}^N \sum_{j=1}^N {\phi}_{r,i} {\phi}^*_{r,j}\, [{\bf R}]_{i,j}\, [{\bf H}_r]_{:,i}[{\bf H}_r]^H_{:,j}.
\end{eqnarray}
We can compute the trace operator of the matrix $\overline{\bf R}_r$ in \eqref{eq:Rr} as
\begin{eqnarray}\label{eq:trRr}
    {\rm tr} \left(\overline{\bf R}_r\right) &=& {\rm tr} \left(\sum_{i=1}^N \sum_{j=1}^N {\phi}_{r,i} {\phi}^*_{r,j}\, [{\bf R}]_{i,j}\, [{\bf H}_r]_{:,i}[{\bf H}_r]^H_{:,j}\right),\nonumber\\
    &=& \sum_{i=1}^N \sum_{j=1}^N {\phi}_{r,i} {\phi}^*_{r,j}\, [{\bf R}]_{i,j}\, {\rm tr} \left([{\bf H}_r]_{:,i}[{\bf H}_r]^H_{:,j}\right),\nonumber\\
    &=& \sum_{i=1}^N \sum_{j=1}^N {\phi}_{r,i} {\phi}^*_{r,j}\, [{\bf R}]_{i,j}\, [{\bf H}_r]^H_{:,j} [{\bf H}_r]_{:,i},\nonumber\\
    &=& \sum_{i=1}^N \sum_{j=1}^N {\phi}^*_{r,j}\,[{\bf H}_r]^H_{:,j} [{\bf R}]_{i,j}\, [{\bf H}_r]_{:,i} \,{\phi}_{r,i}.\nonumber\\
    &=& \sum_{i=1}^N \sum_{j=1}^N {\phi}^*_{r,j}\, [{\bf D}_r]_{j,i}\,{\phi}_{r,i},\nonumber\\
    &=& \boldsymbol{\phi}_r^H {\bf D}_r \boldsymbol{\phi}_r.
\end{eqnarray}
To construct the matrix ${\bf D}_r$, we have that 
\begin{eqnarray}\label{eq:Dr}
    [{\bf D}_r]_{j,i} &=& [{\bf H}_r]^H_{:,j} [{\bf R}]_{i,j}\, [{\bf H}_r]_{:,i},\nonumber\\
    &=& [{\bf H}_r]^H_{:,j} \, [{\bf H}_r]_{:,i} [{\bf R}]_{i,j},\nonumber\\
    &=& \left[{\bf H}_r^H \, {\bf H}_r\right]_{j,i} [{\bf R}]_{i,j},\nonumber\\
    &=& \left[{\bf H}_r^H \, {\bf H}_r\right]_{j,i} [{\bf R}^*]_{j,i},
\end{eqnarray}
in which from the third to the fourth line, we have used the property of the matrix ${\bf R}$ being symmetric (Hermitian, {\it i.e.}, {$[{\bf R}]_{i,j}= [{\bf R}]_{j,i}^*$}, and the main diagonal values are real). Therefore, we can see from \eqref{eq:Dr} that ${\bf D}_r = \left[{\bf H}_r^H \, {\bf H}_r\right] \odot {\bf R}^*$, completing the proof.

\section{Minimizing the result in Equation \eqref{eq:crossProd}}\label{app:AppendB}

{Substituting the expressions of ${\bf R}_k$ in \eqref{eq:Rk2} in \eqref{eq:crossProd}, we obtain}
\begin{equation}
    {\rm tr} \left( {\bf R}_k {\bf R}_{k'} \right) = \beta^{\textsc{ru}}_{r,k} \beta^{\textsc{ru}}_{r',k'} \, \left( T_1 + T_2 + T_3 + T_4 \right),
\end{equation}
{in which}
\begin{align}\label{eq:crossProd2}
    T_1 &= {\rm tr} \left( {\bf H}_r \, {\bf \underline{R}}_r\, {\bf H}_r^H {\bf H}_{r'} \, {\bf \underline{R}}_{r'}^H {\bf H}_{r'}^H \right)\\
    T_2 &= {\rm tr} \left( {\bf H}_r \, {\bf \underline{R}}_r {\bf H}_r^H {\bf R}^{\textsc{bu}}_{k'} \right),\nonumber\\
    T_3 &= {\rm tr} \left( {\bf R}^{\textsc{bu}}_{k} {\bf H}_{r'} \, {\bf \underline{R}}_{r'}^H {\bf H}_{r'}^H \right),\nonumber\\
    T_4 &= {\rm tr} \left( {\bf R}^{\textsc{bu}}_{k} {\bf R}^{\textsc{bu}}_{k'} \right),\nonumber
\end{align}
\normalsize
{with ${\bf \underline{R}}_r = {\rm diag}(\boldsymbol{\phi}_r) {\bf R}\, {\rm diag}(\boldsymbol{\phi}_r)^H$. It is difficult to find a solution to ensure ${\rm tr} \left( {\bf R}_k {\bf R}_{k'} \right) = 0$, since for example $T_4$ does not depend {on} any RIS parameter. However, a simple condition for reducing $T_1$ can be obtained, which depends on the angle of deployment of the RIS. For this sake, we rewrite $T_1$ in \eqref{eq:crossProd2} as}
\begin{align}\label{eq:crossProd3}
    T_1 &= {\rm tr} \left( {\bf H}_{r'}^H {\bf H}_r \, {\bf \underline{R}}_r\, {\bf H}_r^H {\bf H}_{r'} \, {\bf \underline{R}}_{r'}^H \right).
\end{align}

{Then, considering only the LoS part of ${\bf H}_{r}$ and ${\bf H}_{r'}$, \emph{i.e.}, approximating them as ${\bf H}_{r} = \sqrt{\frac{\beta^{\textsc{br}}_r \alpha^{\textsc{br}}_r}{1+\alpha^{\textsc{br}}_r}} \bm{a}_{\textsc{b}}(\vartheta^{\textsc{b}}_r) \,\bm{a}_{\textsc{r}}(\vartheta^{\textsc{r}}_r,\varphi^{\textsc{r}}_r)^H$ and ${\bf H}_{r'} = \sqrt{\frac{\beta^{\textsc{br}}_{r'} \alpha^{\textsc{br}}_{r'}}{1+\alpha^{\textsc{br}}_{r'}}} \bm{a}_{\textsc{b}}(\vartheta^{\textsc{b}}_{r'}) \,\bm{a}_{\textsc{r}}(\vartheta^{\textsc{r}}_{r'},\varphi^{\textsc{r}}_{r'})^H$, we observe that ${\bf H}_{r'}^H {\bf H}_r$ is given by}
\begin{align}\label{eq:crossProd4}
    {\bf H}_{r'}^H {\bf H}_r &\approx \mu_{r,r'} \bm{a}_{\textsc{r}}(\vartheta^{\textsc{r}}_{r'},\varphi^{\textsc{r}}_{r'}) \, \bm{a}_{\textsc{b}}(\vartheta^{\textsc{b}}_{r'})^H \bm{a}_{\textsc{b}}(\vartheta^{\textsc{b}}_r) \,\bm{a}_{\textsc{r}}(\vartheta^{\textsc{r}}_r,\varphi^{\textsc{r}}_r)^H,
\end{align}
{with $\mu_{r,r'} = \sqrt{\frac{\beta^{\textsc{br}}_{r'} \alpha^{\textsc{br}}_{r'}\beta^{\textsc{br}}_r \alpha^{\textsc{br}}_r}{(1+\alpha^{\textsc{br}}_{r'})(1+\alpha^{\textsc{br}}_{r})}}$. Thus one can see that the result in \eqref{eq:crossProd4} is equal to 0 if $\bm{a}_{\textsc{b}}(\vartheta^{\textsc{b}}_{r'})^H \bm{a}_{\textsc{b}}(\vartheta^{\textsc{b}}_r) = 0$, justifying our proposed methodology for the RIS deployment positions. Although the method does not ensure a strict null result for ${\rm tr} \left( {\bf R}_k {\bf R}_{k'} \right)$, our numerical results show that the interference levels can be remarkably reduced. Besides, one can expect that ${\rm tr} \left( {\bf R}_k {\bf R}_{k'} \right) \to 0$ as long as the BS-UEs direct links become severely attenuated and the BS-RIS links become predominantly LoS.}



\vfill

\end{document}